\documentclass[twocolumn,trackchanges]{aastex63}

\usepackage{longtable}
\usepackage{graphicx}

\usepackage{savesym}
\savesymbol{tablenum}
\usepackage{siunitx}
\restoresymbol{SIX}{tablenum}

\usepackage{makecell}
\usepackage{multirow}
\newcommand{\project}[1]{\textsl{#1}}                                           
\newcommand{\JWST}{\project{JWST}}        
\newcommand{\HST}{\project{HST}}          
\newcommand{\Spitzer}{\project{Spitzer}}  
\newcommand{\TESS}{\project{TESS}}     
\newcommand{\CHEOPS}{\project{CHEOPS}}
\newcommand{\bjdtdb}{${\rm {BJD_{TDB}}}$}

\colorlet{Mycolor1}{green!72!red!228!}

\shorttitle{Atmospheric Characterization of KELT-11b}
\shortauthors{Col\'{o}n et al.}

\begin{document}

\title{An Unusual Transmission Spectrum for the Sub-Saturn KELT-11b Suggestive of a Sub-Solar Water Abundance} 

\correspondingauthor{Knicole D. Col\'{o}n}
\email{knicole.colon@nasa.gov}

\author[0000-0001-8020-7121]{Knicole D. Col\'{o}n}
\affiliation{NASA Goddard Space Flight Center, Exoplanets and Stellar Astrophysics Laboratory (Code 667), Greenbelt, MD 20771, USA}		
\affiliation{GSFC Sellers Exoplanet Environments Collaboration}
	
\author[0000-0003-0514-1147]{Laura Kreidberg}
\affiliation{Center for Astrophysics \textbar \ Harvard \& Smithsonian, 60 Garden St, Cambridge, MA 02138, USA}
\affiliation{Max Planck Institute for Astronomy, K{\"o}nigstuhl 17, 69117 Heidelberg, Germany}

\author[0000-0003-0156-4564]{Luis Welbanks}
\affiliation{Institute of Astronomy, University of Cambridge, Madingley Road Cambridge CB3 0HA, UK}

\author[0000-0002-2338-476X]{Michael R. Line}
\affiliation{School of Earth \& Space Exploration, Arizona State University, Tempe, AZ 85257, USA }

\author[0000-0002-4869-000X]{Nikku Madhusudhan}
\affiliation{Institute of Astronomy, University of Cambridge, Madingley Road Cambridge CB3 0HA, UK}

\author[0000-0002-9539-4203]{Thomas Beatty}
\affiliation{Department of Astronomy and Steward Observatory, University of Arizona, Tucson, AZ 85721, USA}

\author[0000-0003-2171-5083]{Patrick Tamburo}
\affiliation{Department of Astronomy \& The Institute for Astrophysical Research, Boston University, 725 Commonwealth Ave., Boston, MA 02215, USA}

\author[0000-0002-7352-7941]{Kevin B. Stevenson}
\affiliation{Johns Hopkins APL, 11100 Johns Hopkins Rd, Laurel, MD 20723, USA}

\author[0000-0002-8119-3355]{Avi Mandell}
\affiliation{NASA Goddard Space Flight Center, Exoplanets and Stellar Astrophysics Laboratory (Code 693), Greenbelt, MD 20771, USA}	
\affiliation{GSFC Sellers Exoplanet Environments Collaboration}

\author[0000-0001-8812-0565]{Joseph E. Rodriguez}
\affiliation{Center for Astrophysics \textbar \ Harvard \& Smithsonian, 60 Garden St, Cambridge, MA 02138, USA}

\author[0000-0001-7139-2724]{Thomas Barclay}
\affiliation{NASA Goddard Space Flight Center, Exoplanets and Stellar Astrophysics Laboratory (Code 667), Greenbelt, MD 20771, USA}
\affiliation{University of Maryland, Baltimore County, 1000 Hilltop Cir, Baltimore, MD 21250, USA}
\affiliation{GSFC Sellers Exoplanet Environments Collaboration}

\author{Eric D. Lopez} 
\affiliation{NASA Goddard Space Flight Center, Exoplanets and Stellar Astrophysics Laboratory (Code 693), Greenbelt, MD 20771, USA}	
\affiliation{GSFC Sellers Exoplanet Environments Collaboration}

\author[0000-0002-3481-9052]{Keivan G. Stassun}
\affiliation{Department of Physics \& Astronomy, Vanderbilt University, 6301 Stevenson Center Ln., Nashville, TN 37235, USA}

\author[0000-0001-6138-8633]{Daniel Angerhausen}
\affiliation{Institute for Particle Physics and Astrophysics, ETH Zurich, Wolfgang-Pauli-Strasse 27, 8093 Zurich, Switzerland}
\affiliation{Blue Marble Space Institute of Science, 1001 4th Ave, Suite 3201, Seattle, WA 98154, USA}

\author[0000-0002-9843-4354]{Jonathan J. Fortney}
\affiliation{Department of Astronomy and Astrophysics, University of California, Santa Cruz, CA 95064, USA}

\author[0000-0001-5160-4486]{David J. James}
\affiliation{ASTRAVEO, LLC, PO Box 1668, Gloucester, MA 01931, USA}

\author[0000-0002-3827-8417]{Joshua Pepper}
\affiliation{Department of Physics, Lehigh University, 16 Memorial Drive East, Bethlehem, PA 18015, USA}

\author[0000-0003-2086-7712]{John P. Ahlers}
\altaffiliation{NASA Postdoctoral Program Fellow}
\affiliation{NASA Goddard Space Flight Center, Exoplanets and Stellar Astrophysics Laboratory (Code 667), Greenbelt, MD 20771, USA}	
\affiliation{GSFC Sellers Exoplanet Environments Collaboration}


\author[0000-0002-8864-1667]{Peter Plavchan}
\affiliation{Department of Physics and Astronomy, George Mason University, 4400 University Drive, MSN 3F3, Fairfax, VA 22030, USA}

\author[0000-0003-3251-3583]{Supachai Awiphan}
\affiliation{National Astronomical Research Institute of Thailand, 260 Moo 4, Donkaew, Mae Rim, Chiang Mai, 50180, Thailand}

\author{Cliff Kotnik} 
\affiliation{American Association of Variable Star Observers, USA}

\author[0000-0001-9504-1486]{Kim K. McLeod}
\affiliation{Department of Astronomy, Wellesley College, Wellesley, MA 02481, USA}

\author[0000-0001-7809-1457]{Gabriel Murawski}
\affiliation{Gabriel Murawski Private Observatory (SOTES)}

\author{Heena Chotani}
\affiliation{Department of Physics and Astronomy, George Mason University, 4400 University Drive, MSN 3F3, Fairfax, VA 22030, USA}

\author{Danny LeBrun} 
\affiliation{Department of Physics and Astronomy, George Mason University, 4400 University Drive, MSN 3F3, Fairfax, VA 22030, USA}
\affiliation{Lockheed Martin, Bethesda, MD 20817, USA}

\author[0000-0003-3937-562X]{William Matzko}
\affiliation{Department of Physics and Astronomy, George Mason University, 4400 University Drive, MSN 3F3, Fairfax, VA 22030, USA}

\author{David Rea} 
\affiliation{Department of Physics and Astronomy, Iowa State University, Ames, IA 50011, USA}

\author{Monica Vidaurri}
\affiliation{Department of Physics and Astronomy, George Mason University, 4400 University Drive, MSN 3F3, Fairfax, VA 22030, USA}

\author{Scott Webster}
\affiliation{Department of Physics and Astronomy, George Mason University, 4400 University Drive, MSN 3F3, Fairfax, VA 22030, USA}

\author{James K. Williams} 
\affiliation{Department of Physics and Astronomy, George Mason University, 4400 University Drive, MSN 3F3, Fairfax, VA 22030, USA}

\author[0000-0001-7526-5116]{Leafia Sheraden Cox} 
\affiliation{Department of Astronomy, Wellesley College, Wellesley, MA 02481, USA}

\author{Nicole Tan} 
\affiliation{Department of Astronomy, Wellesley College, Wellesley, MA 02481, USA}

\author[0000-0002-0388-8004]{Emily A. Gilbert}
\affiliation{Department of Astronomy and Astrophysics, University of
Chicago, 5640 S. Ellis Ave, Chicago, IL 60637, USA}
\affiliation{University of Maryland, Baltimore County, 1000 Hilltop Circle, Baltimore, MD 21250, USA}
\affiliation{The Adler Planetarium, 1300 South Lakeshore Drive, Chicago, IL 60605, USA}
\affiliation{NASA Goddard Space Flight Center, Exoplanets and Stellar Astrophysics Laboratory (Code 667), Greenbelt, MD 20771, USA}		
\affiliation{GSFC Sellers Exoplanet Environments Collaboration}

\begin{abstract}

We present an optical-to-infrared transmission spectrum of the inflated sub-Saturn KELT-11b measured with the Transiting Exoplanet Survey Satellite (\TESS), the Hubble Space Telescope (\HST) Wide Field Camera 3 G141 spectroscopic grism, and the Spitzer Space Telescope (\Spitzer) at 3.6 $\mu$m, in addition to a \Spitzer~4.5 $\mu$m secondary eclipse.  The precise \HST~transmission spectrum notably reveals a low-amplitude water feature with an unusual shape. Based on free retrieval analyses with varying molecular abundances, we find strong evidence for water absorption. Depending on model assumptions, we also find tentative evidence for other absorbers (HCN, TiO, and AlO). The retrieved water abundance is generally $\lesssim 0.1\times$ solar (0.001--0.7$\times$ solar over a range of model assumptions), several orders of magnitude lower than expected from planet formation models based on the solar system metallicity trend. We also consider chemical equilibrium and self-consistent 1D radiative-convective equilibrium model fits and find they too prefer low metallicities ($[M/H] \lesssim -2$, consistent with the free retrieval results). However, all the retrievals should be interpreted with some caution since they either require additional absorbers that are far out of chemical equilibrium to explain the shape of the spectrum or are simply poor fits to the data. Finally, we find the \Spitzer~secondary eclipse is indicative of full heat redistribution from KELT-11b's dayside to nightside, assuming a clear dayside. These potentially unusual results for KELT-11b's composition are suggestive of new challenges on the horizon for atmosphere and formation models in the face of increasingly precise measurements of exoplanet spectra.

\end{abstract}


\section{Introduction}
\label{intro}

In recent years, there has been an explosion of exoplanet atmosphere characterization efforts using both ground- and space-based facilities. The Hubble Space Telescope (\HST) has been especially key to providing a glimpse into the composition of exoplanet atmospheres. \HST~has been used extensively to look for water in particular in the atmospheres of a diverse group of exoplanets, ranging from super-Earths to hot Jupiters \citep[e.g.,][]{sing2016Natur.529...59S,crossfield2017AJ....154..261C,fu17,pinhas2019,welbanks2019ApJ...887L..20W}. In some cases, it has been possible to determine water abundances based on the detection of water absorption features from \HST~\citep[e.g.,][]{kreidberg2014,madhusudhan2014,wakeford2018}. Because water is expected to be the dominant component by mass of icy planetesimals in solar composition protoplanetary disks, measuring the water abundance provides the opportunity to test predictions of core accretion models of planet formation \citep[e.g.,][]{madhusudhan2014,marboeuf2014,lee2016}.

Here, we present an investigation to search for water and other species in the atmosphere of KELT-11b based on observations from the \HST/Wide Field Camera 3 (WFC3), the \Spitzer/Infrared Array Camera (IRAC), and the Transiting Exoplanet Survey Satellite (\TESS). KELT-11b has a mass of just 0.171$\pm$0.015 $M_J$ and a radius of 1.35$\pm$0.10 $R_J$, making it extremely inflated and giving it one of the lowest surface gravities of any planet discovered to date \citep{beatty2017,pepper2017}. With a period of 4.74 days, KELT-11b also has a high equilibrium temperature (1712$^{+51}_{-46}$ K) as reported by \citet{pepper2017} along with a very bright host star ($V$ = 8.0, $K$ = 6.1). Furthermore, its host star is a metal-rich sub-giant ([Fe/H] = 0.17$\pm$0.07; log $g_{\star}$ = 3.7$\pm$0.1) that is part of the ``Retired A-star'' class. Altogether, KELT-11b is one of the best and most interesting targets for atmospheric characterization.

KELT-11b is notably part of an emerging population of low surface gravity sub-Saturn-mass exoplanets (hereafter, ``inflated sub-Saturns'') that are ideal targets for atmospheric characterization via transmission spectroscopy. Other notable planets in this population include WASP-39b \citep{faedi2011}, WASP-107b \citep{anderson2017}, WASP-127b \citep{lam2017}, and HAT-P-67b \citep{zhou2017AJ....153..211Z}. These planets occupy a relatively unexplored corner of parameter space that presents a test for models of planet formation. Planets in this key transitional population likely have similar formation mechanisms as inflated super-Earths \citep[``super-puffs'';][]{lee2016}, such that they formed via run-away core accretion near the snow-line before migrating inward. Such planets are predicted to have very high atmospheric water abundances \citep{lee2016}. 
 
The atmospheres of several of the inflated sub-Saturns noted above have been studied in some detail already. A complete transmission spectrum of WASP-39b collected over 0.3--1.7 $\mu$m revealed a very metal-rich atmosphere ($\sim$150$\times$~solar) with absorption signatures from water (H$_2$O) as well as the alkali metals sodium (Na) and potassium (K) \citep{wakeford2018,kirk2019AJ....158..144K}. However, \citet{pinhas2019} found a slightly sub-solar atmospheric metallicity for WASP-39b, while \citet{welbanks2019ApJ...887L..20W} do support super-solar abundances albeit find one case where the abundance falls below the expected metallicity (based on the trend of what is seen in the solar system for CH$_4$ abundances). The H$_2$O abundance in WASP-107b's atmosphere is consistent with a solar composition and metallicity ($<$30$\times$~solar), although it may be depleted in methane (CH$_4$) relative to solar abundances \citep{kreidberg2018}. Intriguingly, helium (He) has also been found to be escaping from WASP-107b's atmosphere \citep{spake2018,allart2019,kirk2020AJ....159..115K}. Lastly, WASP-127b displays absorption signatures from Na, lithium (Li), K, H$_2$O, and carbon dioxide (CO$_2$), along with evidence for a haze. Correspondingly, WASP-127b has been found to have a moderately metal-rich atmosphere ($\sim$30$\times$~solar) although estimates of the atmospheric metallicity also extend down to sub-solar values \citep{chen2018,spake2019arXiv191108859S,welbanks2019ApJ...887L..20W}. 

The only atmospheric study of KELT-11b conducted previously was in the optical, which revealed no sign of Na absorption in its atmosphere and hinted at the presence of high, thick clouds \citep{zak2019AJ....158..120Z}. This result stands out since other inflated sub-Saturns appear to have at least one common atmospheric trait so far -- feature-rich atmospheres that are not hidden by significant hazes or clouds. To add to this sample of well-characterized inflated exoplanets, we undertook an investigation of the atmosphere of the inflated sub-Saturn KELT-11b using observations from \HST, \Spitzer, and \TESS. We present in this work the results from our investigation, along with ground-based observations used to monitor the activity of KELT-11 around the time of the \HST~transit observations and \Spitzer~secondary eclipse observations. We describe our observations, data reduction, and light curve analyses in Sections \ref{obs} and \ref{lc_fits}, our atmospheric retrievals in Section \ref{atmos}, and wrap up with a discussion and summary in Sections \ref{discussion} and \ref{summary}. 

\section{Observations and Data Reduction}
\label{obs}

A summary of the observations presented here is given in Table \ref{tab:observations}. The observations include a transit of KELT-11b from \HST, a transit and a secondary eclipse from \Spitzer, five full (and one partial) transits from \TESS, and baseline observations from four different ground-based facilities. These observations are described in further detail in the following sections.

\begin{table}[]
\begin{footnotesize}
\begin{centering}
\begin{tabular}{llll}
Facility & Date (UT) & Event & Bandpass \\
\hline
\hline
\Spitzer/IRAC  & 2016 Apr 4 & Transit & 3.6 $\mu$m  \\
\HST/WFC3  & 2018 Apr 28 & Transit & 1.1--1.7 $\mu$m  \\
\TESS  & 2019 Feb 28--Mar 26 & Transit & 0.6--1.0 $\mu$m  \\
\Spitzer/IRAC  & 2018 Apr 11 & Eclipse & 4.5 $\mu$m  \\
GMU & 2018 Apr 11 & Baseline & $I$ \\
GMU & 2018 Apr 12 & Baseline & $I$ \\
GMU & 2018 Apr 19 & Baseline & $I$ \\
TRT-TNO & 2018 Apr 12 & Baseline & $R$ \\
TRT-TNO & 2018 Apr 13 & Baseline & $R$ \\
Pike's Peak & 2018 Apr 15 & Baseline & $Ic$ \\
Wellesley College & 2018 Apr 19 & Baseline & $r$ \\
\end{tabular}
\caption{A summary of the observations of KELT-11b presented in this paper. GMU is George Mason University and TRT-TNO is Thai Robotic Telescope-Thai National Observatory. The \TESS~observations covered five full transits of KELT-11b and one partial transit.}
\label{tab:observations}
\end{centering}
\end{footnotesize}
\end{table}

\subsection{HST/WFC3}
\label{obs_hst}
We observed a single transit of KELT-11b with \HST/WFC3 on UT 2018 April 18 between 0410 UT and 1735 UT (\HST~Program GO 15255; Co-PIs K. Col\'on and L. Kreidberg).  Our observations spanned 9 \HST~orbits, and at the beginning of each orbit we obtained a direct image of the target star with the F130N filter. The remaining exposures used the G141 filter and employed the spatial scan observing mode, with a scan rate of 0.96 arcsec/second. We used the readout mode \texttt{SPARS\_25} with \texttt{NSAMP} = 3, yielding an exposure time of 46.696 seconds. This observing setup yielded a spatial scan 340 pixels long, so we used the $512\times512$ subarray to ensure that we captured the entire spectrum. The peak photoelectron count per exposure was $4.9\times10^4$. We generally obtained 16 exposures per orbit, with the exception of orbits 4, 8, and 9, which were trimmed by a few minutes to allow for a gyro bias update and crossing the South Atlantic Anomaly. For these orbits, we added several additional direct images to the beginning of the observation to fill up the buffer and force a buffer dump after the orbit ended. 

For our primary data reduction, we used custom software described in detail by \cite{kreidberg2014}. We extracted each up-the-ramp sample separately, using an extraction window that extended 250 pixels in the spatial direction. To estimate the background counts, we identified a region of the image that was uncontaminated by flux from KELT-11 or any background stars. We took the median of photoelectron counts in this rectangular region spanning rows $6-50$ and columns $6 - 30$.  We subtracted the background, optimally extracted the spectrum from each up-the-ramp sample \citep{horne86},  co-added the samples, and summed in the spatial direction to obtain a final spectrum from each exposure. There is minimal spectral drift over the course of the observation (typically 5 angstroms per orbit and 10 angstroms over the entire 9-orbit visit).

As a means to validate our results, we performed a secondary reduction of the data using custom software described by \citet{Stevenson2014a}.  For each up-the-ramp sample, we estimated the background on a column-by-column basis using regions above and below the spectral extraction region, whose optimal height was determined to be 204 pixels in the spatial direction.   We used optimal spectral extraction on each sample, aligned the 1D spectra along the dispersion direction, and co-added the reads from each exposure to obtain a time series of 1D spectra. Our analysis of both of these reductions is discussed below in Section \ref{lc_fits}.

\subsection{Spitzer}
\label{obs_spitzer}

A single transit of KELT-11b was previously observed with \Spitzer/IRAC Channel 1 (3.6 $\mu$m) starting on UT 2016 April 4 as part of \Spitzer~Program GO 12096 (PI: T. Beatty). These observations and subsequent analysis are described in detail in \cite{beatty2017}. Here, we present a new analysis of the \Spitzer~transit data that we performed in order to derive a more robust measurement of the transit depth at 3.6 $\mu$m to use in combination with the \HST~and \TESS~transit data for KELT-11b. The new analysis is described further in Section \ref{spitzer_transit_fits} below. 

Approximately one week prior to the \HST~transit observations of KELT-11b, we observed a single secondary eclipse of KELT-11b with \Spitzer/IRAC Channel 2 (4.5 $\mu$m) as part of \Spitzer~Program GO 13229 (PI: K. Col\'on). These occurred from UT 2018 April 11 0141 and UT 2018 April 11 1631. For the eclipse observations we used subarray mode with 0.1 second exposures, and PCRS peak-up mode with KELT-11 as the peak-up target to stabilize the spacecraft's pointing. We initially observed KELT-11 for 0.5 hours to allow the telescope's pointing to settle before beginning the science observations. We discarded these initial settling observations and did not use them in our analysis. In total, we collected 373,925 images at 4.5 $\mu$m.

We began our data reduction and photometric extraction process from the basic calibrated data (BCD) images. The reduction of the KELT-11 images and the extraction of the photometry followed the process in \cite{beatty2019}, and we briefly re-describe it here. We first determined the time of each exposure by assuming that the exposures within an individual 64-image data cube began at the \textsc{mjd\_obs} header time, and were evenly spaced between the \textsc{aintbeg} and \textsc{atimeend} header times. We converted the resulting mid-exposure times to \bjdtdb. 

We next estimated the background level in each image and measured KELT-11's position. We began by masking out a box 15 pixels on a side centered on KELT-11, and taking the median of the unmasked pixels as the background level. To increase the accuracy of our background measurement, we corrected bad pixels and cosmic ray hits by performing an iterative $5\,\sigma$ clipping on the time-series for each individual pixel and replacing outliers with the time-series' median. The average background in our observations was $2.6\,\mathrm{e}^-\,\mathrm{pix}^{-1}$, which was 0.02\% of KELT-11's average flux. We then used the background-subtracted, bad-pixel corrected images to measure the pixel position of KELT-11 in each image using a two-dimensional Gaussian. Note that we used these corrected images only to estimate the background and to measure the position of KELT-11 -- we used uncorrected background-subtracted images for the photometric extraction.

We extracted raw photometry for KELT-11 using a circular extraction aperture centered on KELT-11's position in each image. We found that using a variable aperture radius at 2.4$\times$ the full-width half-maximum of KELT-11 in each image provided the cleanest photometry. For reference, the average full-width half-maximum of KELT-11's point spread function was 2.08 pixels. We tested a range of fixed aperture sizes from 3.0 to 4.5 pixels in radius, but in all cases the log-likelihoods of the resulting best fits were lower, the scatter in the residuals higher, and eclipse properties were consistent with our variable aperture.

Finally, we trimmed outliers from the raw photometry. The first 25 minutes of the eclipse observations showed a clear residual ramp effect, so we excluded the first 15,000 points. We removed outliers from the remaining photometry by fitting a line between the average flux of the first 100 and last 100 points in the remaing data, and clipping those points that were more than $5\,\sigma$ away from that line. We determined the error on each point by adding in quadrature the Poisson noise from KELT-11's flux and the integrated background flux in the photometric aperture. All together, this left us with 373,907 flux measurements at 4.5 $\mu$m. Our analysis of the \Spitzer~4.5 $\mu$m light curve is described in Section \ref{spitzer_eclipse_fits}.

\subsection{TESS}
\label{obs_tess}

KELT-11 (TIC 55092869) was observed at 2-minute cadence\footnote{KELT-11 was included on the 2-minute cadence target list as part of multiple \TESS~Guest Investigator (GI) programs: 11025 (PI: Travis Metcalfe), 11048 (PI: Daniel Huber), 11112 (PI: John Southworth), 11183 (PI: Stephen Kane).} by \TESS~for approximately 27 days in Sector 9 (UT 2019 Feb 28 to UT 2019 Mar 26). Five complete transits and one partial transit of KELT-11b were observed in that time. \TESS~has a single broad optical bandpass, spanning from 600--1000 nm. All \TESS~data are calibrated by the Science Processing Operations Center (SPOC) at NASA Ames Research Center. For each 2-minute cadence target, the SPOC generates systematic error-corrected light curves using an optimal photometric aperture \citep{jenkins2016}. This light curve has passed through the pre-search data conditioning (PDC) module of the \TESS~pipeline \citep{smith2012kepler, stumpe2014multiscale, jenkins2016}. This version of the \TESS~light curve has also been corrected for instrumental signals and contaminating light from nearby stars. The SPOC light curve for KELT-11 was downloaded from the Mikulski Archive for Space Telescopes (MAST) to be used in our analysis, which is described in Section \ref{tess_transit_fits}.

\subsection{Ground-Based Observatories}
\label{obs_ground}

\begin{figure*}[ht]
\begin{center}
\includegraphics[scale = 0.5]{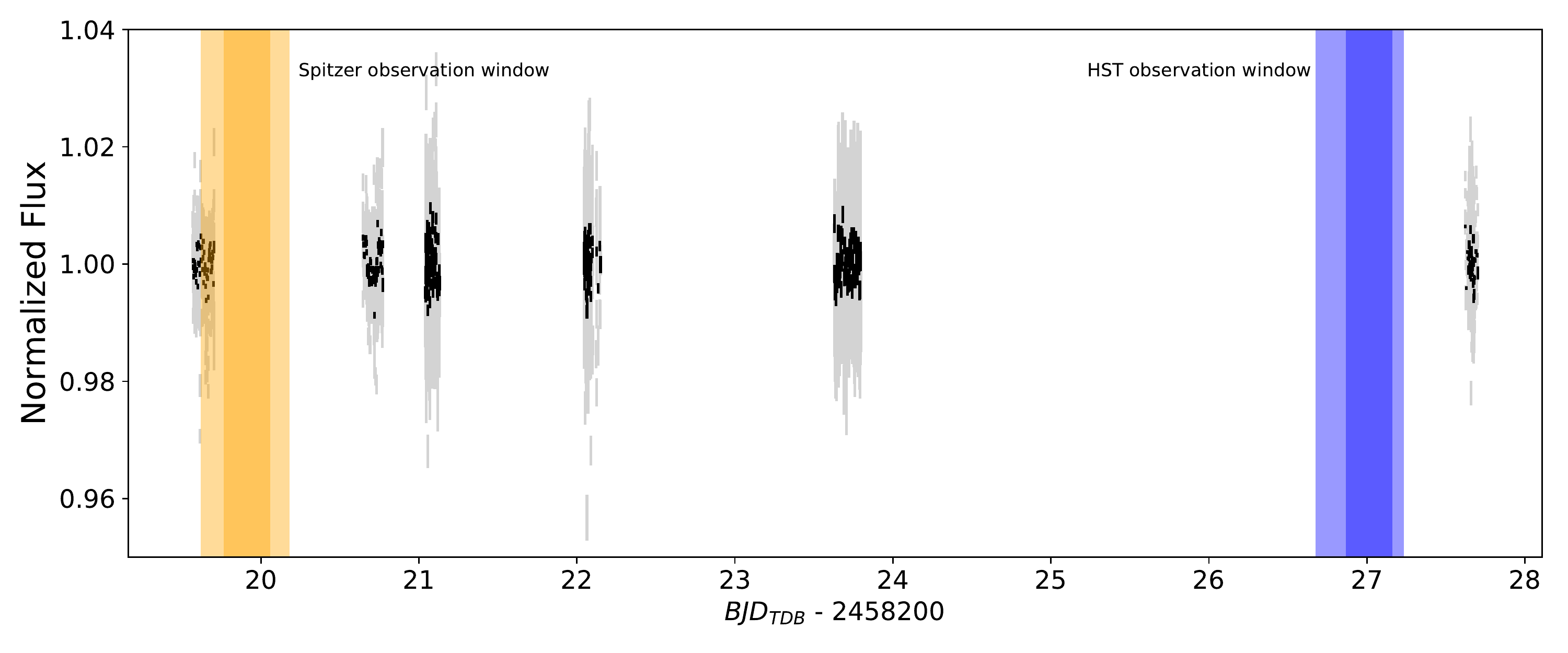}
\caption{Ground-based light curves of KELT-11, which were collected from four different observatories around the times of the \Spitzer~eclipse and \HST~transit observations of KELT-11b. The light gray points are unbinned data and black points are binned data. The window of the \Spitzer~observations is shown in orange, with the time of the eclipse highlighted in a darker shade of orange. Similarly, the window of the \HST~observations is shown in blue, with the time of the transit highlighted in a darker shade of blue.}
\label{fig:all_ground}
\end{center}
\end{figure*}

\begin{figure}
\begin{center}
\includegraphics[scale = 0.7]{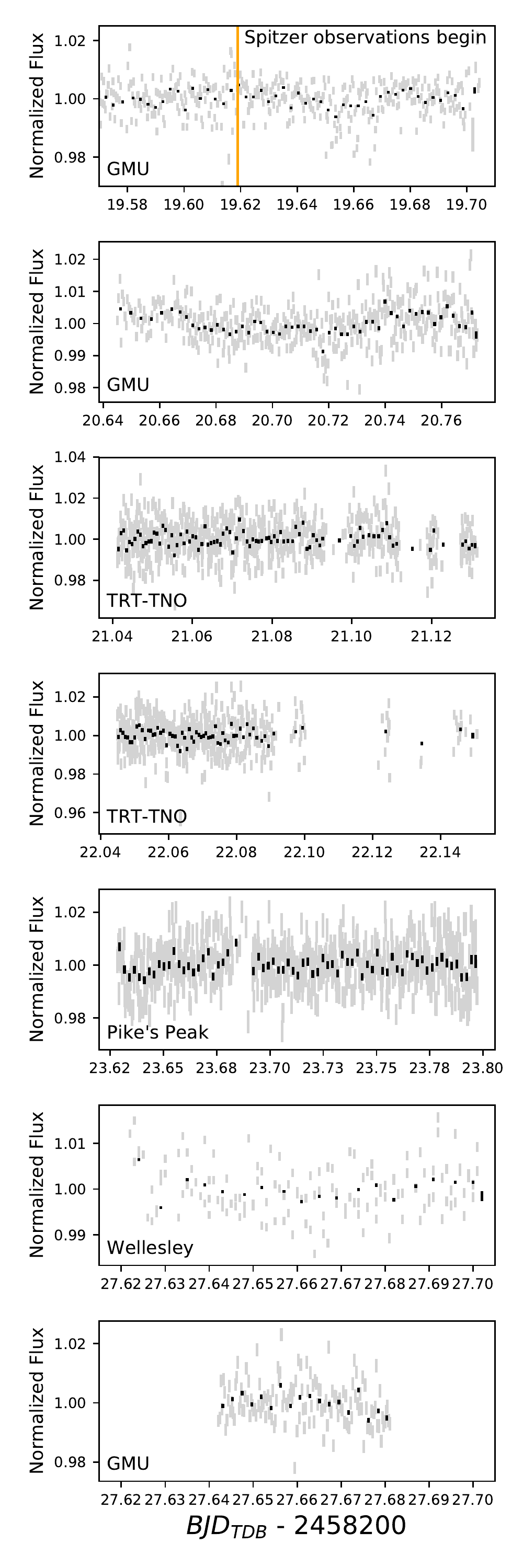}
\caption{Individual ground-based light curves of KELT-11. Colors are the same as in Figure \ref{fig:all_ground}. The time that \Spitzer~baseline observations began is noted in the top panel. The eclipse of KELT-11b began after the GMU observations had already ended.}
\label{fig:each_ground}
\end{center}
\end{figure}

To monitor stellar activity around the time of the \Spitzer~eclipse and \HST~transit observations on UT 2018 April 11 and 18, respectively, we collected a series of optical observations with ground-based telescopes. These observations were planned using the TAPIR-based \citep{jensen2013ascl.soft06007J} Kilodegree Extremely Little Telescope (KELT) Transit Finder (KTF) web tool. The full time coverage of the ground-based observations is shown in Figure \ref{fig:all_ground}, with the windows of the \Spitzer~eclipse and \HST~transit observations marked for reference. Individual ground-based light curves are shown in Figure \ref{fig:each_ground}.

We collected observations of KELT-11 in $I$-band on UT 2018 April 11, 12, and 19 with the 0.8 meter George Mason University (GMU) telescope. The observations on UT 2018 April 11 used an exposure time of 15 sec and spanned 3.26 hours. These observations overlapped with part of the baseline \Spitzer~observations collected prior to the eclipse (Figures \ref{fig:all_ground} and \ref{fig:each_ground}). KELT-11 was also monitored after the \Spitzer~eclipse observations, on UT 2018 April 12 for 3.06 hours, with an exposure time of 10 sec. On UT 2018 April 19, KELT-11 was observed for 0.934 hours with an exposure time of 10 sec. These observations took place after the \HST~transit observations.

KELT-11 was monitored in $R$-band for 2.16 and 2.54 hours on UT 2018 April 12 and 13, respectively, with the 0.5 meter Thai Robotic
Telescope—Thai National Observatory (TRT-TNO) in Chiang Mai, Thailand. An exposure time of 2 sec was used, resulting in 1043 and 948 images on the two nights. A few hundred saturated images were removed prior to analysis. Both of these data sets were collected after the \Spitzer~eclipse observations.

We observed KELT-11 in the $Ic$-band for 4.05 hours on UT 2018 April 15 with the 0.36 meter telescope at Pike's Peak, Colorado. The exposure time was 10 sec. These observations took place approximately mid-way between the \Spitzer~eclipse and \HST~transit observations. 

Lastly, KELT-11 was observed on UT 2018 April 19 in $r$-band using the 0.6096 meter telescope at Wellesley College (Whitin Observatory). An exposure time of 24 sec was used to observe KELT-11 for a duration of 1.92 hours. These observations overlapped with the last set of GMU observations, both of which took place after the \HST~transit observations.

All data were calibrated by the respective observers and simple aperture photometry was performed using AstroImageJ \citep{collins2017}. To account for systematic effects, most of the data sets were detrended against one or more different parameters, e.g., airmass, the full-width half-maximum of the target star, the centroid measurements of the target position on the detector, or the total counts of the comparison star ensemble.

\section{Light Curve Analysis}
\label{lc_fits}

In the following sections, we describe our analysis of the \HST, \Spitzer, and \TESS~transit light curves and the \Spitzer~eclipse light curve. In Section \ref{updates} we include a summary of the updated parameters for KELT-11b we find based on these light curve analyses.

\subsection{HST/WFC3 White Light Transit}
\label{hst_white_fits}
To create a broadband, ``white" light curve from the \HST\ data, we summed up each spectrum over the full length of the spectral trace. We fit the light curve with a joint model of the transit and instrument systematics commonly seen in WFC3 time series observations \citep{zhou17}. To model the transit, we used the \texttt{batman} package \citep{kreidberg15}. We fit for the following free parameters: the planet-to-star radius ratio $R_p/R_s$, the orbital inclination $i$, the ratio of semi-major axis to stellar radius $a/R_s$, the time of inferior conjunction $T_c$, and a linear limb-darkening parameter $u_1$. We fixed the orbital period on the best fit from the discovery paper, 4.7365 days \citep{pepper2017}. We also fit a model to the instrument systematics that included a quadratic visit-long trend and exponential orbit-long trends, as well as a constant offset between forward and reverse scan directions. This systematics model is identical to Equation 3 of \cite{kreidberg18b}. This functional form for the systematics is dubbed the \texttt{model-ramp} technique \citep{kreidberg2014}. We followed common practice and dropped the first orbit of the visit, which had larger systematic noise.

\begin{figure}[h!]
\begin{center}
\includegraphics[scale = 0.8]{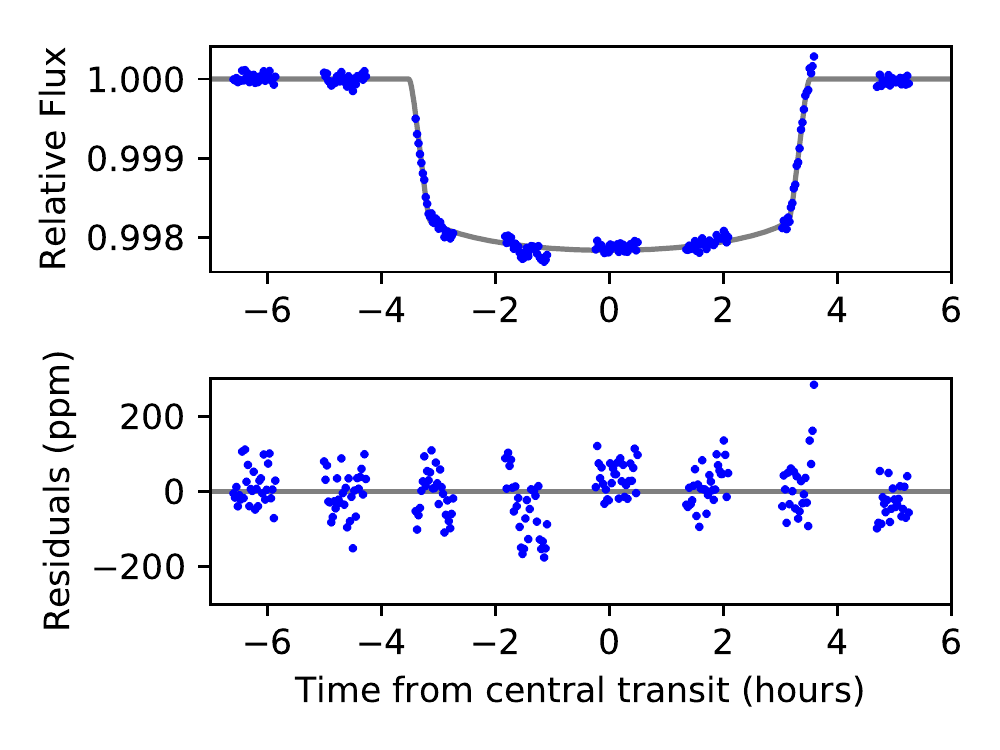}
\caption{\HST~white light transit of KELT-11b. The top panel shows the normalized, systematics-removed data (blue points) compared to the best fit transit model (black line). The bottom panel shows the residuals from the best fit (points) compared to a horizontal line to guide the eye.}
\label{fig:hst_white}
\end{center}
\end{figure}

Figure \ref{fig:hst_white} presents the \HST/WFC3 G141 white light transit of KELT-11b. The best fit transit light curve has a root-mean-square (rms) variability of 65 parts per million (ppm). This is roughly $3\times$ higher than the expected photon noise (22 ppm), but is comparable to the best achieved white light precision from WFC3 spatial scanning observations of bright stars \citep{knutson14}.  To account for this additional scatter, we increased the per point error by a constant scale factor so that the best fit light curve model had a reduced $\chi^2$ of unity. We ran a Markov chain Monte Carlo fit to the light curve with the \texttt{emcee} package \citep{foremanmackey13}. The fit was initialized with the best fit parameters. The chain had 10,000 steps and 50 walkers, and we discarded the first 20\% of the chain as burn-in. The resulting transit parameters are listed in Table\,\ref{tab:white_lc_params}.

\begin{table}[]
\begin{tabular}{ll}
Parameter                  & Value\\
\hline
\hline
$R_p/R_s$                  & $0.045182  \pm 0.00028$ \\
$T_c$ (BJD$_\mathrm{TDB}$) & $2458227.01450 \pm 0.00012$          \\
$a/R_s$                    & $4.83 \pm 0.09$                      \\
$i$ (degrees)              & $84.4 \pm 0.5$                       \\
$u_1$                      & $0.271 \pm 0.015$    
\end{tabular}
\caption{Transit parameters from the \HST/WFC3 broadband light curve fit. The values are the medians and 68\% credible intervals. The derived transit depth $(R_{p}/R_{s})^2$ = $2041\pm25$ ppm or $0.2041\pm0.0025 \%$.}
\label{tab:white_lc_params}
\end{table}

\begin{table}[]
\begin{centering}
\begin{tabular}{llll}
Wavelength ($\mu$m) & ($R_p/R_s)$$^2$ (\%) & $u_1$ & $\chi^2_\nu$\\
\hline
\hline
$ 1.125 - 1.150 $ & $ 0.2031 \pm \num{ 1.8e-03 } $ & $ 0.312 \pm 0.018 $ & 1.04 \\
$ 1.150 - 1.175 $ & $ 0.1994 \pm \num{ 1.7e-03 } $ & $ 0.317 \pm 0.019 $ & 1.00 \\
$ 1.175 - 1.200 $ & $ 0.1982 \pm \num{ 1.7e-03 } $ & $ 0.301 \pm 0.019 $ & 1.07 \\
$ 1.200 - 1.225 $ & $ 0.1985 \pm \num{ 1.6e-03 } $ & $ 0.291 \pm 0.018 $ & 0.81 \\
$ 1.225 - 1.250 $ & $ 0.1963 \pm \num{ 1.5e-03 } $ & $ 0.310 \pm 0.018 $ & 0.81 \\
$ 1.250 - 1.275 $ & $ 0.1977 \pm \num{ 1.5e-03 } $ & $ 0.294 \pm 0.018 $ & 0.95 \\
$ 1.275 - 1.300 $ & $ 0.1989 \pm \num{ 1.5e-03 } $ & $ 0.268 \pm 0.017 $ & 0.88 \\
$ 1.300 - 1.325 $ & $ 0.1928 \pm \num{ 1.5e-03 } $ & $ 0.293 \pm 0.018 $ & 0.90 \\
$ 1.325 - 1.350 $ & $ 0.1961 \pm \num{ 1.5e-03 } $ & $ 0.301 \pm 0.017 $ & 0.83 \\
$ 1.350 - 1.375 $ & $ 0.2044 \pm \num{ 1.5e-03 } $ & $ 0.308 \pm 0.017 $ & 1.01 \\
$ 1.375 - 1.400 $ & $ 0.2068 \pm \num{ 1.5e-03 } $ & $ 0.298 \pm 0.017 $ & 0.92 \\
$ 1.400 - 1.425 $ & $ 0.2070 \pm \num{ 1.4e-03 } $ & $ 0.225 \pm 0.017 $ & 0.86 \\
$ 1.425 - 1.450 $ & $ 0.2107 \pm \num{ 1.5e-03 } $ & $ 0.231 \pm 0.018 $ & 1.01 \\
$ 1.450 - 1.475 $ & $ 0.2093 \pm \num{ 1.6e-03 } $ & $ 0.228 \pm 0.018 $ & 0.89 \\
$ 1.475 - 1.500 $ & $ 0.2097 \pm \num{ 1.7e-03 } $ & $ 0.253 \pm 0.018 $ & 1.08 \\
$ 1.500 - 1.525 $ & $ 0.2053 \pm \num{ 1.8e-03 } $ & $ 0.244 \pm 0.018 $ & 1.15 \\
$ 1.525 - 1.550 $ & $ 0.2067 \pm \num{ 1.7e-03 } $ & $ 0.224 \pm 0.019 $ & 0.95 \\
$ 1.550 - 1.575 $ & $ 0.2072 \pm \num{ 1.7e-03 } $ & $ 0.250 \pm 0.019 $ & 0.91 \\
$ 1.575 - 1.600 $ & $ 0.2082 \pm \num{ 1.9e-03 } $ & $ 0.213 \pm 0.020 $ & 1.35 \\
$ 1.600 - 1.625 $ & $ 0.2052 \pm \num{ 2.4e-03 } $ & $ 0.191 \pm 0.020 $ & 1.91 \\
$ 1.625 - 1.650 $ & $ 0.2028 \pm \num{ 2.9e-03 } $ & $ 0.218 \pm 0.021 $ & 2.71 \\
\end{tabular}
\caption{Transit depths and linear limb-darkening parameter $u_1$ from the spectroscopic light curve MCMC fits to the \HST/WFC3 data. The values are the median and 68\% credible interval from the posterior distributions. The reduced $\chi^2$ values for the best fit are listed in the right-most column. For light curves with $\chi^2_\nu > 1$, the per point uncertainties were scaled up in the MCMC to yield $\chi^2_\nu = 1$.}
\label{tab:spec_lc_params}
\end{centering}
\end{table}

\subsection{HST/WFC3 Spectroscopic Transit}
\label{hst_spec_fits}
To generate spectroscopic light curves, we binned the spectrum into 21 spectrophotometric channels in the wavelength range 1.125--1.65 $\mu$m. We fit each spectroscopic light curve with a joint transit and systematics model. The transit time was fixed to the best fit value from the white light curve fit ($T_c = 2458227.01455$ BJD$_\mathrm{TDB}$). 

The transit model allowed $R_p/R_s$ and $u_1$ to vary, but fixed the orbital parameters $a/R_s = 5.00$ and $i = 85.3$ to be consistent with the original parameters measured from the \Spitzer~transit \citep{beatty2017}. We note that in Section \ref{spitzer_transit_fits} we present a new analysis of the \Spitzer~transit data, and in Section \ref{tess_transit_fits} we present analyses of the \TESS~transit data. The results from these new analyses are consistent with the original parameters derived in \citet{beatty2017} to better than 2$\sigma$.

\begin{figure}[h!]
\begin{center}
\includegraphics[scale = 0.7]{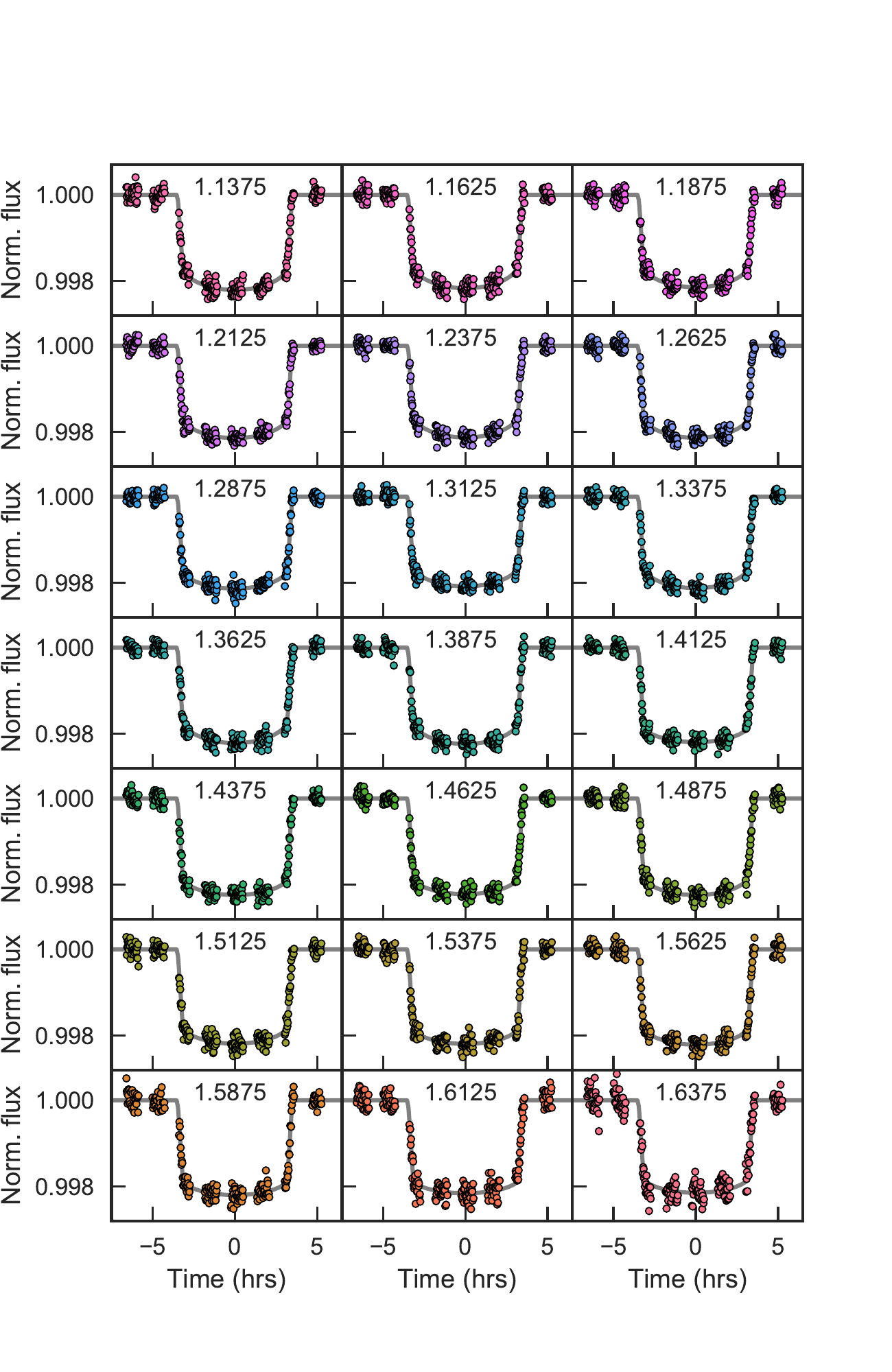}
\caption{\HST~spectroscopic transit light curves (points). The light curves are normalized to unity and corrected for systematics. The best fit transit models are also shown (gray lines). Each panel is annotated with the central wavelength of the spectroscopic channel in microns.}
\label{fig:wfc3_spec}
\end{center}
\end{figure}

To fit the systematics, we used the \texttt{divide-white} technique, which scales the residuals from the best fit white light curve \citep{stevenson14, kreidberg2014}. We also fit a linear, visit-long slope to each spectral channel. The free parameters for the systematics model were the constant scale factor for the white light residuals and a linear slope. For each spectroscopic light curve, we ran an MCMC using the same approach described in Section \ref{hst_white_fits}. Table\,\ref{tab:spec_lc_params} lists the resulting transit depth, limb-darkening value, and $\chi^2$ for each fit. For light curves where the best fit had a reduced chi-squared ($\chi^2_\nu$) greater than unity, we rescaled the per point uncertainties to achieve $\chi^2_\nu = 1$ before we ran the MCMC. The best fit normalized spectroscopic light curves are shown in Figure\,\ref{fig:wfc3_spec}. 

\begin{figure*}
\begin{center}
    \includegraphics[scale = 0.6]{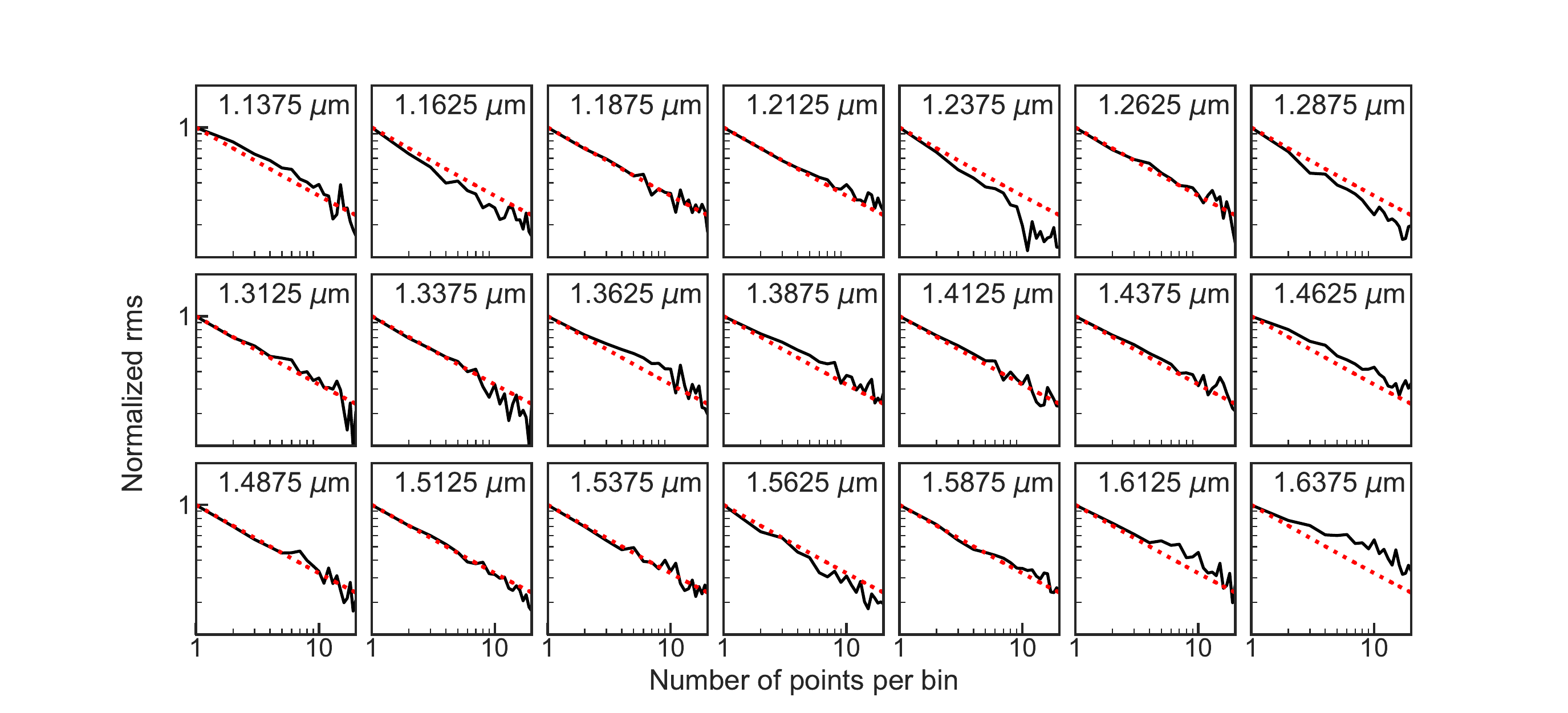}
\end{center}
\caption{The root mean square deviation for the WFC3 spectroscopic light curves as a function of points per bin (black lines). The expected trend for Poisson noise is indicated by the dashed red line. The wavelength of the spectroscopic channel is noted at the top of each panel.}
\label{fig:allan}
\end{figure*}

To evaluate whether correlated noise is present in the spectroscopic light curve, we computed the rms of the light curve residuals over a range of bin sizes for each spectroscopic channel (shown in Figure\,\ref{fig:allan}). The rms deviation decreases proportionally to the square root of the number of points per bin ($\sqrt{N}$), as expected for Poisson noise that is uncorrelated in time. This indicates that the correlated noise visible in the residuals to the white light curve fit (Figure~\ref{fig:hst_white}) are effectively removed from the spectroscopic light curves with the \texttt{divide-white} technique.

\subsubsection{Comparison with Alternate Systematics Models}
Because the KELT-11b spectrum is one of the most precise WFC3 transmission spectra ever published \citep{knutson14, Line2016}, we compared several different models for the instrument systematics to ensure that our results are not biased by our choice of systematics model. We considered two models in addition to the \texttt{divide-white} model: (1) the analytic \texttt{model-ramp} function used to fit the white light curve (described in Section \ref{hst_white_fits}), and (2) the \texttt{RECTE} model from \cite{zhou17} that uses data from the first orbit rather than discarding it. We also compared the results to an independent analysis from co-author K.~B.~Stevenson (discussed below). Figure \ref{fig:hst_model_comparison} compares results from four different analyses of the \HST/WFC3 data. The spectroscopic transit depths typically agree to much better than $1\sigma$, modulo a constant offset. The mean $(R_p/R_s)^2$ values for the different model fits differ by up to 0.036\%; however, this does not affect the atmospheric retrieval because the retrieval marginalizes over the uncertainty in the planet radius.  We chose the \texttt{divide-white} model to use in our subsequent atmospheric retrieval because it has the fewest free parameters and lowest rms for the best fit light curve fits.

\subsubsection{Comparison with Independent Pipeline Fit}
\begin{figure}[t]
\begin{center}
\includegraphics[width = 0.49\textwidth]{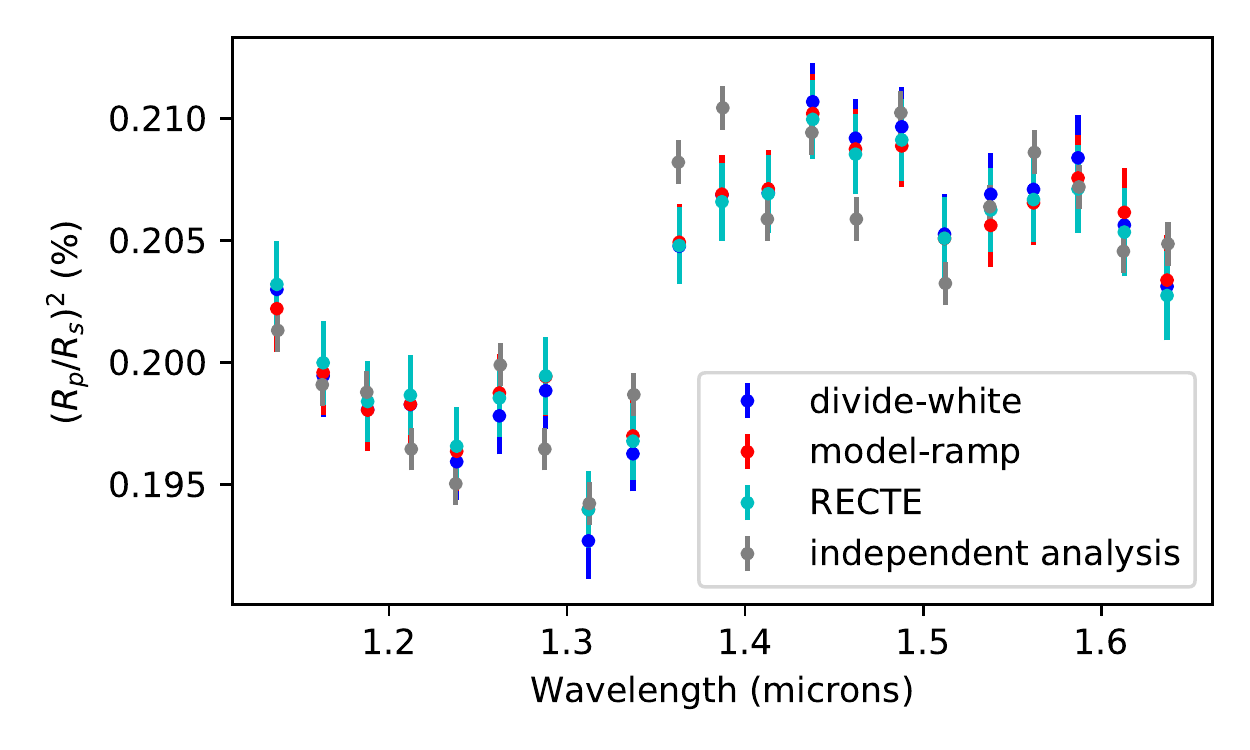}
\caption{Comparison of the \HST/WFC3~transmission spectrum of KELT-11b derived with different analysis techniques. The spectra have a constant offset applied so they have the same mean value, allowing a comparison between the shapes of the spectra.  The dark blue points are from the \texttt{divide-white} technique, which assumes the instrument systematics are independent of wavelength. We use the \texttt{divide-white} spectrum for the atmospheric retrieval. We also show results for the analytic \texttt{model-ramp} technique (red points) and the physically-motivated \texttt{RECTE} model from \cite{zhou17} (cyan points).  The results from the independent analysis of K.~B.~ Stevenson are shown in gray.}
\label{fig:hst_model_comparison}
\end{center}
\end{figure}

The independent analysis by Stevenson also used the \texttt{divide-white} method when fitting the spectroscopic light curves.  For each spectroscopic channel, we fit a transit model, a linear ramp in time, and a flux offset between the forward and reverse scans.  We used the Exoplanet Characterization ToolKit (ExoCTK)\footnote{https://exoctk.stsci.edu/} to derive fixed quadratic limb-darkening parameters.
In Figure \ref{fig:hst_model_comparison} we show the transmission spectrum of KELT-11b derived from the independent reductions of the \HST/WFC3 data.

\subsection{Spitzer 3.6 $\mu$m Transit}
\label{spitzer_transit_fits}

\Spitzer/IRAC observations at 3.6 and 4.5 $\mu$m are subject to the so-called ``pixel phase" effect, where variations in the sensitivity of different pixels, coupled with pointing jitter, introduce intensity fluctuations to the photometry. These intensity fluctuations can be on the order of 1\% \citep{tamburo2018}, and they must be removed to accurately measure transit and eclipse depths from IRAC data. 

To account for the pixel phase effect for the 3.6 $\mu$m transit of KELT-11b, we utilized the Pixel Level Decorrelation (PLD) algorithm developed by \citet{deming2015}. Briefly, PLD works by fitting the time series intensities of individual pixels, a temporal baseline function, and a transit model to the photometry via linear regression. The best-fit linear regression model is then used to initialize a Markov chain Monte Carlo (MCMC) simulation, which allows for a robust estimate of the uncertainties on physical parameters. 

Our version of PLD permits the use of up to 25 pixels encompassing the stellar PSF in a 5x5 pixel grid. The choice of different sets of pixels introduces different basis vectors to the linear regression, which changes how the pixel phase effect is removed, which in turn changes the physical parameters determined from the data. To choose the best set of pixels, we followed the approach of \citet{dalba2019}, who used a Bayesian Information Criterion (BIC) analysis to choose among the different pixel grid combinations. We find that a grid using all 25 pixels produces the optimal BIC score. In the same analysis, we tested different temporal baseline functions to remove long-term systematics, finding that linear ramps give better performance compared to quadratic or exponential functions. 

\begin{figure}[h!]
\begin{center}
\includegraphics[scale=0.34,angle=0]{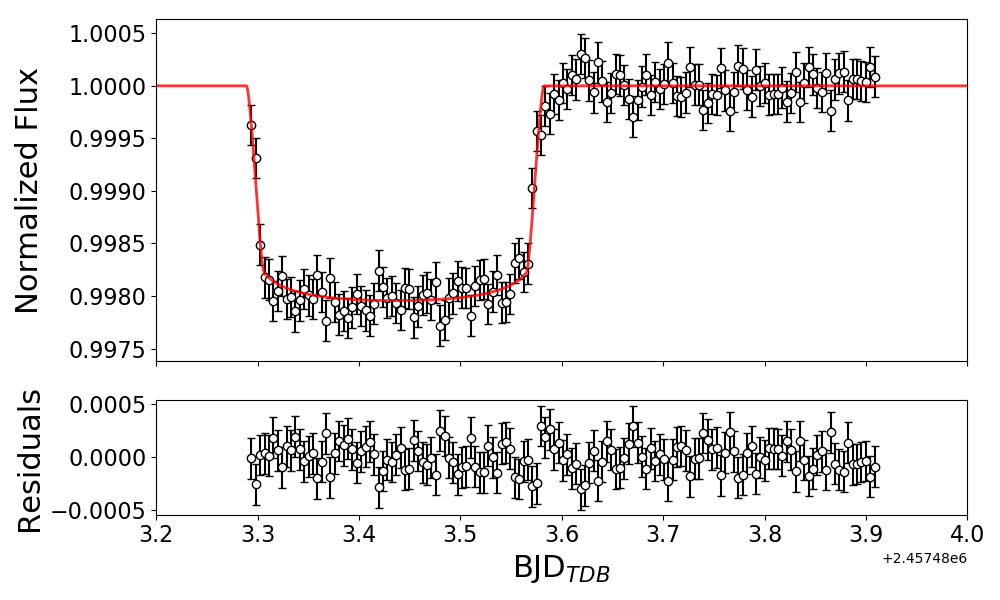}
\caption{\Spitzer~3.6 $\mu$m binned transit light curve of KELT-11b with the best fit model shown in red (top panel). The bottom panel shows the light curve residuals from the best fit.
\label{fig:spitzer_transit}}
\end{center}
\end{figure}

Figure \ref{fig:spitzer_transit} presents our new analysis of the \Spitzer~3.6 $\mu$m transit of KELT-11b, and the transit parameters are given in Table \ref{tab:spitzer_lc_params}. From our PLD analysis and using the \HST~transit time as a prior, we measure $R_p/R_s = $ 0.0443 $\pm$ 0.0011 or a transit depth $(R_p/R_s)^2 = $ 1961 $\pm$ 94 ppm. This is more precise than but still within 1.9$\sigma$ of the value of $R_p/R_s =$ 0.0503 $\pm$ 0.0032 and transit depth $(R_p/R_s)^2 = $2650$^{+350}_{-380}$ ppm measured in \citet{beatty2017} for the same observation, where \citet{beatty2017} used a non-parametric Gaussian Process (GP) regression model to fit the transit \citep{gibson2012}. We note that a residual systematic is seen around the time of transit egress, which was similarly seen in the analysis by \citet{beatty2017}. Since our measurements are consistent with those from \citet{beatty2017}, we conclude this systematic has not affected the measured transit depth. We therefore include the transit depth $(R_p/R_s)^2 = $ 1961 $\pm$ 94 ppm in our analysis of the combined \TESS+\HST+\Spitzer~transmission spectrum.

\begin{table}[]
\begin{tabular}{ll}
Parameter                  & Value\\
\hline
\hline
$R_p/R_s$                  & $0.0443  \pm 0.0011$ \\
$T_c$ (BJD$_\mathrm{TDB}$) & $2457483.4358 \pm 0.0014$          \\
$a/R_s$                    & $4.74 \pm 0.09$                      \\
$i$ (degrees)              & $83.8 \pm 0.5$                       \\
$u_1$                      & $0.070 \pm 0.028$  \\
$u_2$                      & $0.180 \pm 0.035$ \\
\end{tabular}
\caption{Transit parameters from the \textit{Spitzer} 3.6 $\mu$m light curve fit. The values are the medians and 68\% credible intervals. The derived transit depth $(R_{p}/R_{s})^2$ = $1961\pm94$ ppm or $0.1961\pm0.0094 \%$.}
\label{tab:spitzer_lc_params}
\end{table}

\subsection{TESS Optical Transits}
\label{tess_transit_fits}

The transit of KELT-11b has a duration of $\sim$7 hours, and previously published follow-up optical transit photometry of KELT-11b had either been stitched together from multiple ground-based telescopes or collected at high airmass in order to provide full coverage of the transit \citep{beatty2017,pepper2017}. In addition, with a visual magnitude $V$ = 8, there are a limited number of comparably bright comparison stars in the field near KELT-11, which are needed for precise differential photometry from the ground. With data from \TESS~we are able to obtain the first precise optical transit depth measurement for KELT-11b. 

\begin{figure*} 
\begin{center}
\includegraphics[scale=0.65,angle=0]{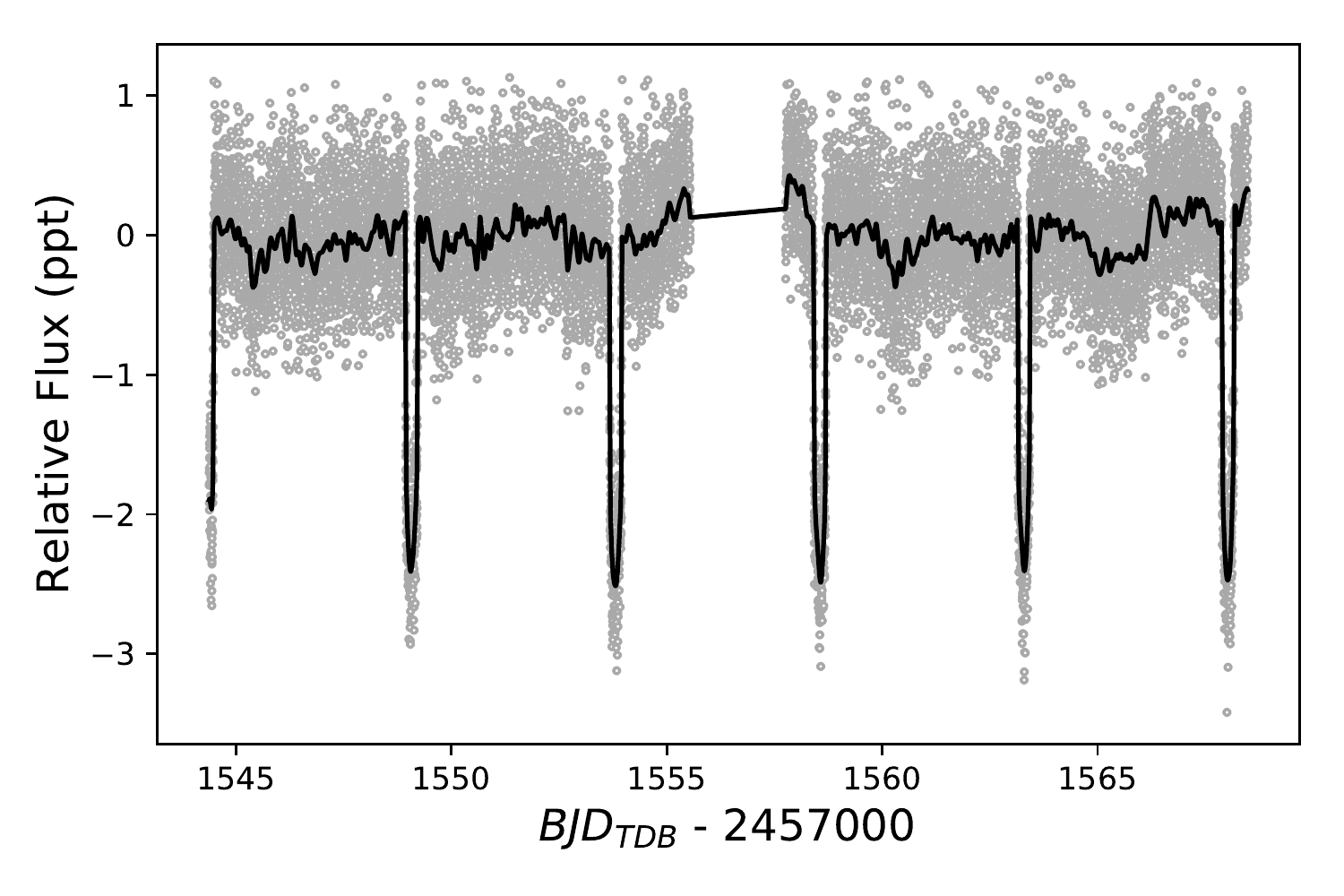}
\includegraphics[scale=0.55,angle=0]{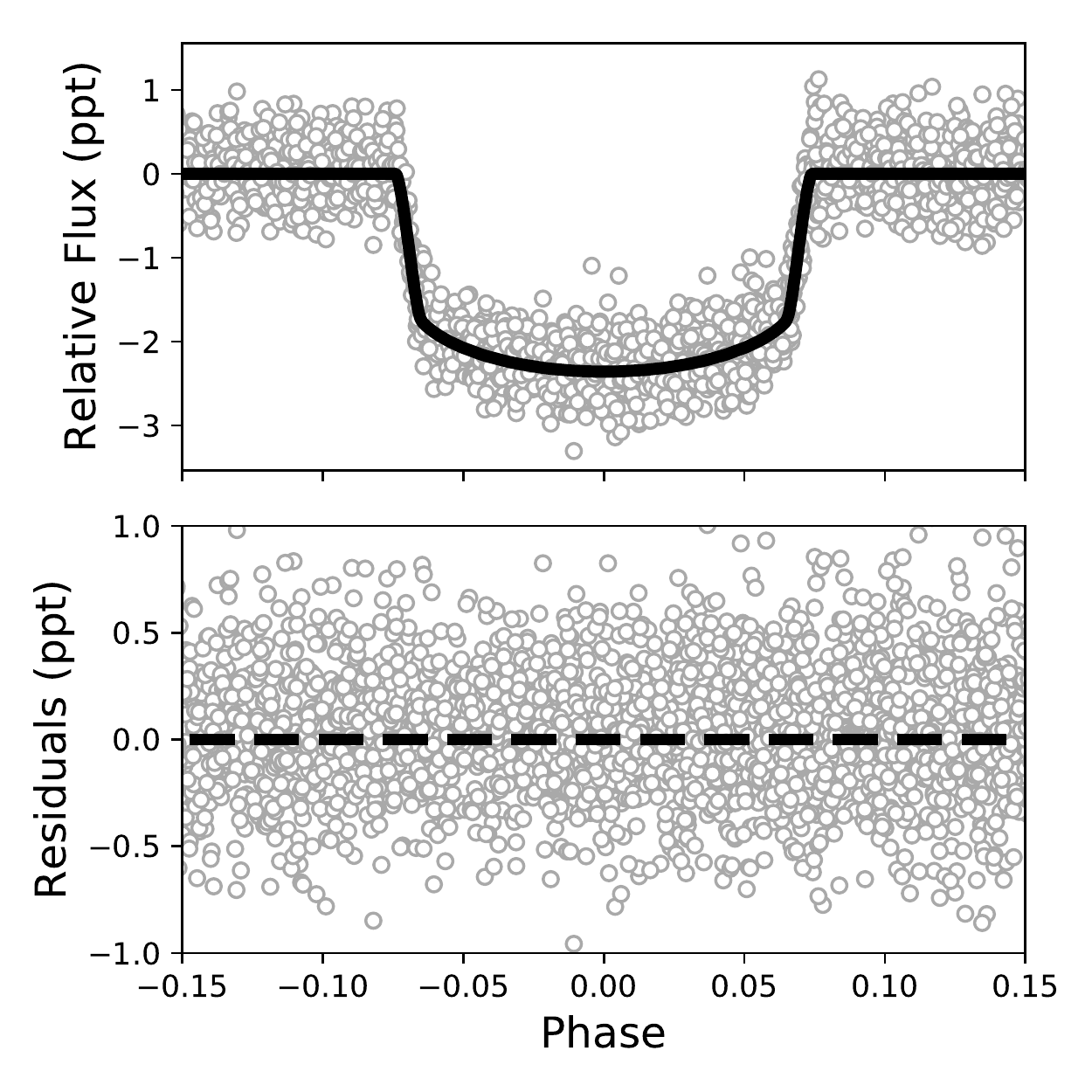}
\caption{\TESS~light curve of KELT-11b. The left panel shows the complete light curve from \TESS~Sector 9 in gray along with the median joint transit light curve and GP model in black. The right panel shows the phase-folded light curve and residuals, after correcting the data against the GP model.
\label{fig:tess}}
\end{center}
\end{figure*}

We used the \texttt{exoplanet} toolkit \citep{exoplanet:exoplanet} to infer the physical properties of KELT-11 from the \TESS~light curve, which is shown in Figure \ref{fig:tess}. The transit model uses the analytic prescription of \citet{exoplanet:agol19} to describe a limb-darkened light curve and is parameterized in the manner recommended by \citet{exoplanet:kipping13}. In addition to limb-darkening parameters, the \texttt{exoplanet} model was parameterized in terms of log stellar density, log orbital period, transit mid-point time, impact parameter, orbital eccentricity, periastron angle and log planet-to-star radius ratio. Correlated noise still present in the PDC light curve -- primarily a mix of uncorrected instrumental signals and any stellar variability -- were modeled as a Gaussian Process (GP) describing a series of stochastically driven simple harmonic oscillators \citep{gibson2012,exoplanet:foremanmackey17,exoplanet:foremanmackey18}. Sampling in \texttt{exoplanet} is built upon the \texttt{pymc3} modeling framework that enabled us to use the highly efficient No U-Turn Sampler \citep{NUTS}. We modeled the log orbital period and transit mid-time as normal distributions with parameters set from the values measured in \citet{beatty2017}. Eccentricity was modeled as a Beta distribution with parameters recommended by \citet{Kipping2013}, and periastron angle was sampled in vector space. The log planet-to-star radius ratio had a broad normal prior with mean of -3.1 and standard deviation of one.

Figure \ref{fig:tess} presents the full and phase-folded \TESS~transit light curve of KELT-11b. The best fit parameters are given in Table~\ref{tab:tess_params} and are consistent with the parameters measured from the fit to the \HST~white light curve. The measured $R_p/R_s = 0.04644\pm0.00065$ corresponds to a transit depth $(R_p/R_s)^2 = 2157\pm60$ ppm in the \TESS~bandpass. This is significantly more precise than the previously reported optical transit depth from \citet{pepper2017} of $(R_p/R_s)^2 = $ $2690_{-260}^{+280}$ ppm. For this reason, we include the \TESS~transit depth in our analysis of the combined \TESS+\HST+\Spitzer~transmission spectrum but not the \citet{pepper2017} transit depth.

We performed an additional analysis of the \TESS~data using \texttt{EXOFASTv2} \citep{eastman2013,eastman2017,eastman2019} to compare to the results derived from \texttt{exoplanet} and to further validate the assumptions made in our analysis of the \HST~and \Spitzer~data. This analysis was performed on the calibrated SPOC light curve, without implementing any additional detrending. As above, we enforced a Gaussian prior on the period and transit time of the planet from \citet{beatty2017}. Allowing the parameters to vary in the \texttt{EXOFASTv2} fit, we find that the derived parameters from \texttt{EXOFASTv2} and \texttt{exoplanet} are typically consistent to well within 1$\sigma$. When we fixed the orbital parameters $a/R_s = 5.00$ and $i = 85.3$ in the \texttt{EXOFASTv2} fit to match the analysis of the \HST~spectroscopic light curves, we find the measured parameters are also consistent with the free fits from both \texttt{EXOFASTv2} and \texttt{exoplanet} and to the \HST~white light curve, validating the various assumptions and results presented here. As the analyses with \texttt{exoplanet} and \texttt{EXOFASTv2} are consistent, we simply choose to adopt the transit depth measured using the \texttt{exoplanet} analysis as the input for the combined \TESS+\HST+\Spitzer~transmission spectrum.

\begin{table}[]
\begin{tabular}{ll}
Parameter                  & Value\\
\hline
\hline
$\rho_*$ (cgs)          & $0.0945_{-0.0076}^{+0.0083}$\\
$P$ (days)              & $4.7362083_{-0.0000040}^{+0.0000041}$\\
$a$ (AU)  	            & $0.0625_{-0.0029}^{+0.0030}$\\
$R_{p}/R_{s}$          & $0.04644\pm0.00065$ \\
$T_c$ (BJD$_\mathrm{TDB}$)     & $2457483.43047_{-0.00081}^{+0.00081}$\\
$a/R_{s}$               & $4.82_{-0.13}^{+0.14}$\\
$i$ (degrees) 	        & $84.18_{-0.64}^{+0.77}$\\
$b$                     & $0.494_{-0.060}^{+0.049}$\\
$T_{dur}$ (hours) 	    & $7.00_{-0.44}^{+0.23}$\\
$eccentricity$ 	                & $0.030_{-0.022}^{+0.038}$\\
$u_0$           & $0.44_{-0.18}^{+0.16}$\\
$u_1$           & $-0.01_{-0.20}^{+0.23}$\\
\end{tabular}
\caption{Median values and 68\% confidence interval for parameters derived from the TESS transit. The derived transit depth $(R_{p}/R_{s})^2$ = $2157\pm60$ ppm or $0.2157\pm0.0060 \%$.}
\label{tab:tess_params}
\end{table}


\subsection{Spitzer 4.5 $\mu$m Secondary Eclipse}
\label{spitzer_eclipse_fits}

We fit the 4.5 $\mu$m secondary eclipse data from \Spitzer~(Section 2.2) using a BLISS mapping analysis \citep{Stevenson2012}. BLISS mapping uses bi-linear interpolation to create a map of the underlying intrapixel sensitivity variations present in the IRAC detectors and is one of the standard techniques for analyzing high precision time series photometry from \Spitzer.

The specific fitting procedure we used was the same as in \cite{beatty2019}, though restricted to only considering an astrophysical model for the eclipse. We constructed the underlying interpolation grid using a spacing of 0.03 pixels in both the x- and y-directions, which was more than three times the standard deviation in the image-to-image changes in KELT-11's location on the detector. We included a background exponential trend with time on top of the BLISS map, to further remove a long term drift in the raw photometry.

We modeled the eclipse itself using a \texttt{batman} model, and we applied a set of Gaussian priors to nearly all of the eclipse parameters -- except for the eclipse depth itself -- based on their values and uncertainties from \cite{beatty2017} and the ephemeris determined from our fit to the \TESS\ data.

We began the BLISS mapping fit to the eclipse by conducting an initial Nelder-Mead likelihood minimization, which we followed by an MCMC run initialized about the Nelder-Mead minimum. We ran the MCMC process for an initial 30,000 step burn-in, followed by a 300,000 step production run. We verified that the production chains were converged by checking that the Gelman-Rubin statistic for each fit parameter was less than 1.1, and by a visual inspection of the MCMC corner plot.

\begin{figure}[h!]
\begin{center}
\includegraphics[scale=0.65,angle=0]{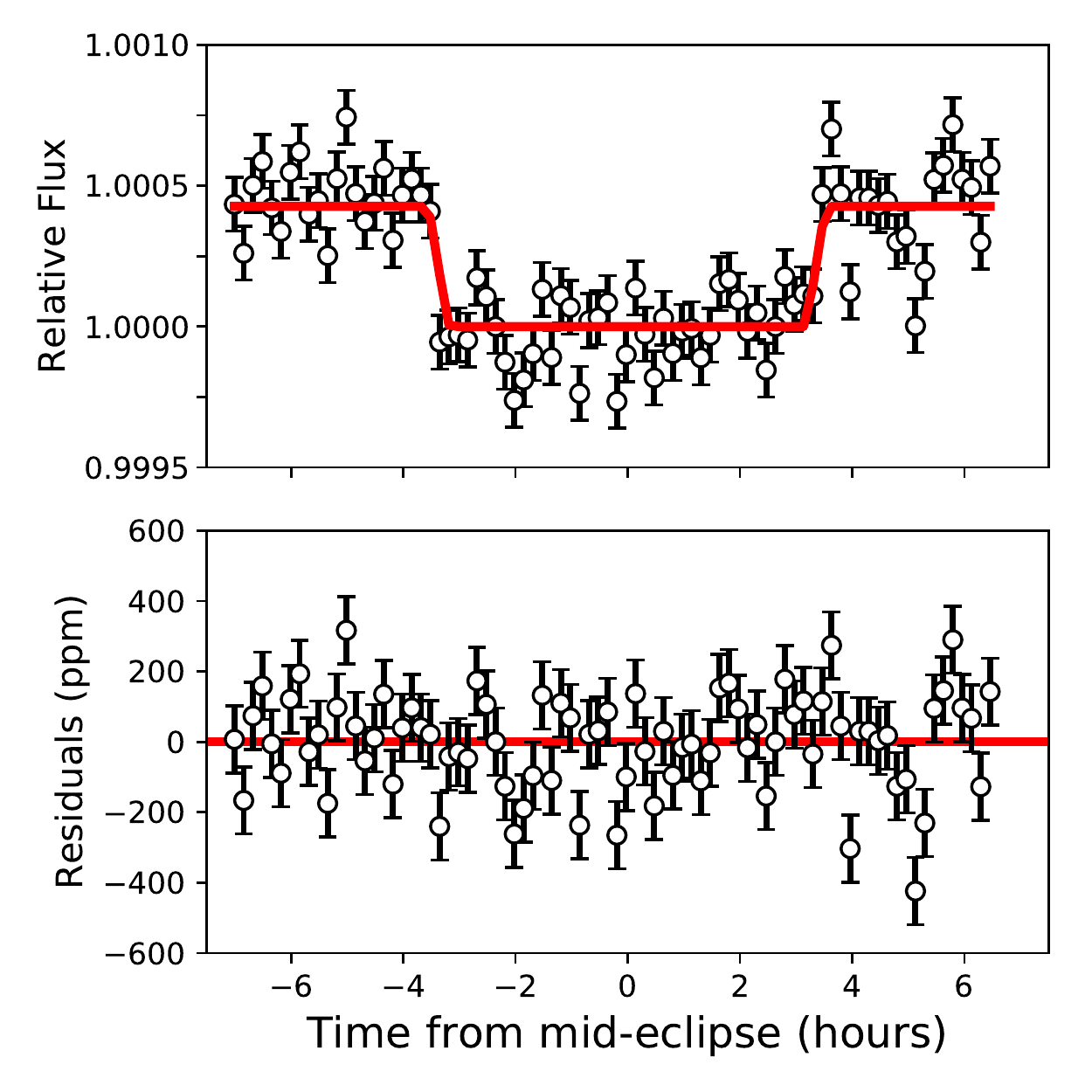}
\caption{\Spitzer~4.5 $\mu$m binned secondary eclipse light curve of KELT-11b. The best fit eclipse model is shown as the solid red line. The bottom panel shows the light curve residuals from the best fit.
\label{fig:spitzer_eclipse}}
\end{center}
\end{figure}

Table \ref{tab:eclipse_params} shows the primary results from our fit to the 4.5 $\mu$m eclipse data. We clearly detect the eclipse, with a depth of 427$\pm$42 ppm (Figure \ref{fig:spitzer_eclipse}). From the eclipse timing and duration the orbit of KELT-11b appears perfectly circular with both $\sqrt{e}\cos\omega$ and  $\sqrt{e}\sin\omega$ measured as zero. The timing of the secondary eclipse occurs exactly half a period after the nearest transit. This agrees with our analysis of the \TESS\ photometry (Section \ref{tess_transit_fits}), where we find an orbital eccentricity of $0.030_{-0.022}^{+0.038}$ (i.e. consistent with an eccentricity of zero). The other five parameters in the eclipse model ($T_C$, $\log P$, $R_p/R_*$, $\cos i$, $\log a/R_*$) were dominated by the Gaussian priors we imposed upon them from the \HST~and \TESS~fits, and so returned results consistent with those data.

\begin{table}[]
\begin{tabular}{ll}
Parameter                  & Value\\
\hline
\hline
Eclipse~Depth (ppm)                  & $427  \pm 42$ \\
$T_s$ (BJD$_\mathrm{TDB}$) & $2458229.3835 \pm 0.0005$          \\
$\sqrt{e}cos{\omega}$                    & $0.000 \pm 0.010$                      \\
$\sqrt{e}sin{\omega}$               & $0.000 \pm 0.016$                       \\
\end{tabular}
\caption{Eclipse depth, time, and orbital shape parameters from the Spitzer 4.5 $\mu$m eclipse fit.}
\label{tab:eclipse_params}
\end{table}

\subsection{Updated Parameters for the KELT-11 System}
\label{updates}

The \Spitzer~transit observations originally published in \citet{beatty2017} were planned and collected based on the ephemeris from \citet{pepper2017}. \citet{beatty2017} found that the \Spitzer~transit occurred 42 minutes earlier than predicted compared to the \citet{pepper2017} ephemeris. \citet{beatty2017} therefore significantly refined the transit ephemeris of KELT-11b, but the impact was that the transit observed by \Spitzer~arrived early enough that no baseline data were collected prior to the transit. As a result, the measurement of the transit depth was impacted by the partial \Spitzer~transit observations. Our re-analysis of the previously published \Spitzer~transit data here made use of a different analysis method and having the \HST~transit time available as a prior. From this, we provide a new precise transit depth measurement at 3.6 $\mu$m of 1961$\pm$94 ppm. 

The new high-precision \HST~and \TESS~data that covered full transits of KELT-11b along with a complete eclipse of KELT-11b with \Spitzer~have allowed us to refine additional key orbital and physical parameters for this planet as well as study its atmospheric properties. Tables \ref{tab:white_lc_params}, \ref{tab:tess_params}, and \ref{tab:eclipse_params} present the transit and eclipse parameters from the fits to the \HST/WFC3 broadband transit light curve, the \TESS~transit light curve, and the \Spitzer~eclipse light curve. From the \TESS~fit, which included a prior on the orbital period from \citet{beatty2017}, we find that the period is $\sim$7 seconds longer and $\sim$10 times more precise than the period measured in \citet{beatty2017}. We also note the nominal transit time from the \TESS~fit is later than the \HST-derived transit time by 58.2 seconds, but they are still consistent to within 1$\sigma$. From the \TESS~fit we obtained a precise optical transit depth measurement of $(R_p/R_s)^2 = 2157\pm60$ ppm. In comparison, the transit depth we measured from the \HST/WFC3 near-infrared broadband light curve is $(R_p/R_s)^2 = $ $2041\pm25$ ppm. Finally, from the \Spitzer~eclipse observations, we determined that KELT-11b's orbit is fully consistent with a circular orbit. This is in agreement with our fit to the \TESS~data, where we measured an eccentricity consistent with zero ($0.030_{-0.022}^{+0.038}$).


\section{Atmospheric Constraints}
\label{atmos}

In the following sections we present results from our atmospheric retrievals of the combined \TESS+\HST+\Spitzer~transmission spectrum as well as the individual \HST~transmission spectrum. We utilize two different retrieval tools, AURA and CHIMERA, to test the robustness of the results against different modeling assumptions. We present an analysis of the emission spectrum based on our single \Spitzer~eclipse for KELT-11b in Section \ref{emission}.

\subsection{Transmission Spectrum Analysis with AURA}
\label{aura}

We analyze the transmission spectra of KELT-11b using an adaptation of the retrieval code AURA \citep{Pinhas2018} as described in \citet{Welbanks2019a}. The code calculates the transit depth of a planet by computing line by line radiative transfer in a transmission geometry. We consider a one-dimensional atmosphere divided into 100 layers uniformly spaced in $\log_{10}$(P) from $10^{-6}$ to $10^2$ bar under hydrostatic equilibrium.

We retrieve the atmospheric properties of KELT-11b employing models with different degrees of complexity. The pressure-temperature (P-T) profile of the atmosphere is retrieved using either a simple isothermal profile or a more robust parametric profile following the prescription in \cite{Madhusudhan2009}. The chemical composition of the atmosphere is retrieved by assuming uniform volume mixing ratios and treating them as free parameters. Our models include the prominent opacity sources expected in the atmospheres of hot Jupiters \citep[e.g.,][]{Madhusudhan2012}: H$_2$ Rayleigh scattering, H$_2$-H$_2$ and H$_2$-He collision induced absorption \citep[CIA;][]{Richard2012}, H$_2$O \citep{Rothman2010}, Na \citep{Allard2019}, K \citep{Allard2016}, CH$_4$ \citep{Yurchenko2014}, NH$_3$ \citep{Yurchenko2011}, HCN \citep{Barber2014}, CO \citep{Rothman2010}, CO$_2$ \citep{Rothman2010}, TiO \citep{Schwenke1998}, AlO \citep{Patrascu2015}, and VO \citep{McKemmish2016}. The opacities for the chemical species are computed following the methods of \citet{Gandhi2017}.

We also consider the possibility of cloudy and hazy atmospheres with inhomogeneous coverage. We allow for the presence of clouds and/or hazes following the parameterization in \citet{Line2016, MacDonald2017}, as employed in \citet{Welbanks2019a}. Atmospheres with non-homogeneous cloud coverage are the result of a linear superposition of a clear atmosphere and an opaque atmosphere due to clouds and/or hazes through the parameter $\bar{\phi}$, corresponding to the fraction of cloud cover at the terminator. The contribution due to hazes is incorporated as $\sigma=a\sigma_0(\lambda/\lambda_0)^\gamma$, a modification to Rayleigh scattering. In this prescription, $\gamma$ is the scattering slope, $a$ is the Rayleigh-enhancement factor, and $\sigma_0$ is the H$_2$ Rayleigh scattering cross-section ($5.31\times10^{-31}$~m$^2$) at the reference wavelength $\lambda_0=350$~nm. We consider the presence of opaque regions of the atmosphere due to clouds through an opaque (gray) cloud deck with cloud-top pressure P$_{\text{cloud}}$.

\subsubsection{Analysis of the Combined \TESS+\HST+\Spitzer~Transmission Spectrum}

For our analysis of the transmission spectrum of KELT-11b, we perform an initial exploratory retrieval considering absorption due to all the species listed above, inhomogeneous clouds and hazes, and a parametric P-T profile. This exploratory retrieval helps indicate the parameters and species that ought to be considered in the fiducial model and helps assess which chemical species may be present in the transmission spectrum of KELT-11b. We opt for this approach to avoid over-fitting the data and including more parameters than there are observations. We determine a fiducial model that considers absorption due to H$_2$O, Na, K, HCN, AlO, and TiO, 6 parameters for the P-T profile, 1 parameter for the reference pressure (P$_{\text{ref}}$) at the radius of the planet $R_{\text{p}}$, and 4 parameters for clouds/hazes. This model with 17 free parameters is used to retrieve the atmospheric properties of KELT-11b using the complete transmission spectrum comprising of the \TESS~optical, \HST/WFC3 near-infrared, and \Spitzer~infrared observations; a total of 23 spectral points. 

\begin{figure*}
\begin{center}
\includegraphics[width=1.\textwidth]{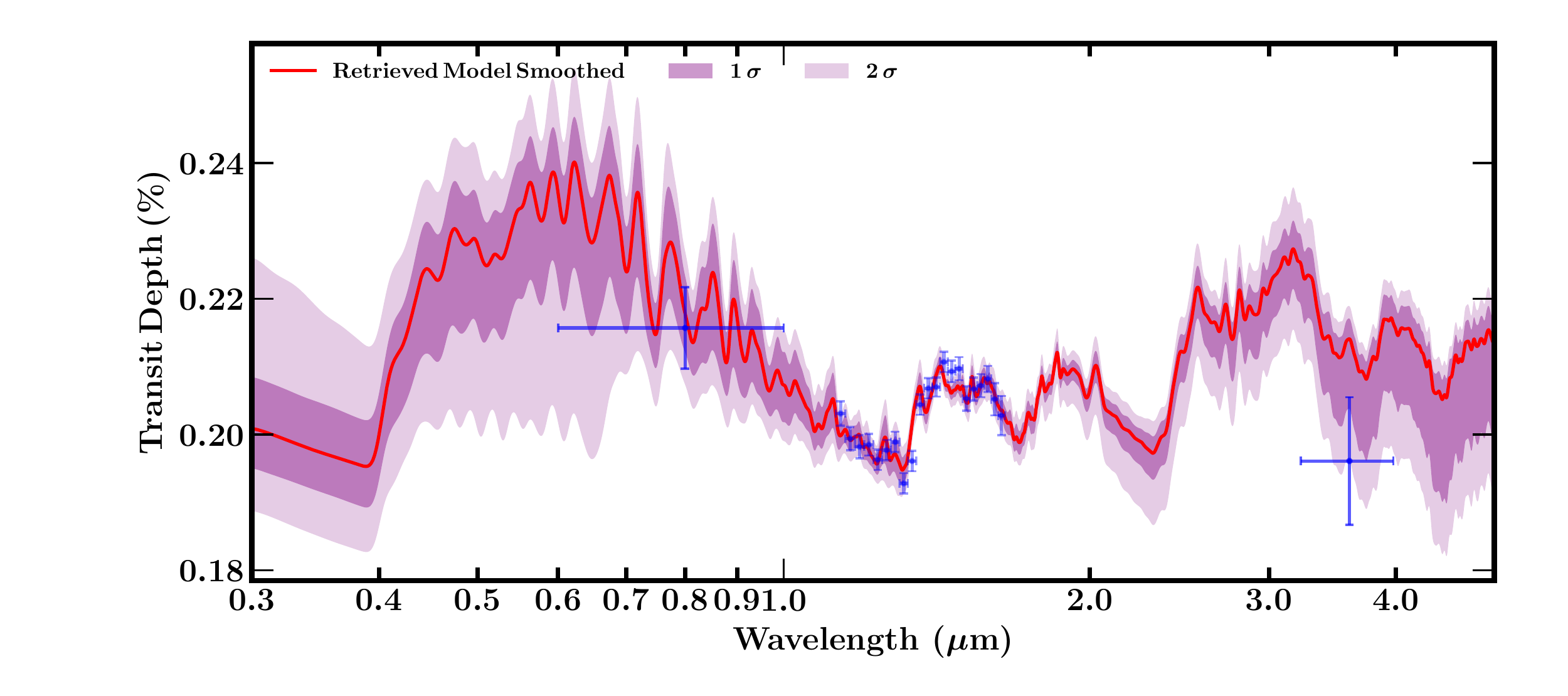}
\caption{Retrieved transmission spectrum of KELT-11b for the fiducial 17 parameter model. The retrieved median transmission spectrum is shown in red with 1$\sigma$ and 2$\sigma$ contours shown in purple shaded regions. \TESS, \HST/WFC3, and \Spitzer~observations are shown using blue markers. The best fit model has a $\chi^{2}$ of  29.97  for  6 degrees of freedom. The p-value is 3.98$\times10^{-05}$ and the BIC is 83.27.
\label{fig:aura_full}}
\end{center}
\end{figure*}

\begin{figure*}[h!]
\begin{center}
\includegraphics[width=1.0\textwidth]{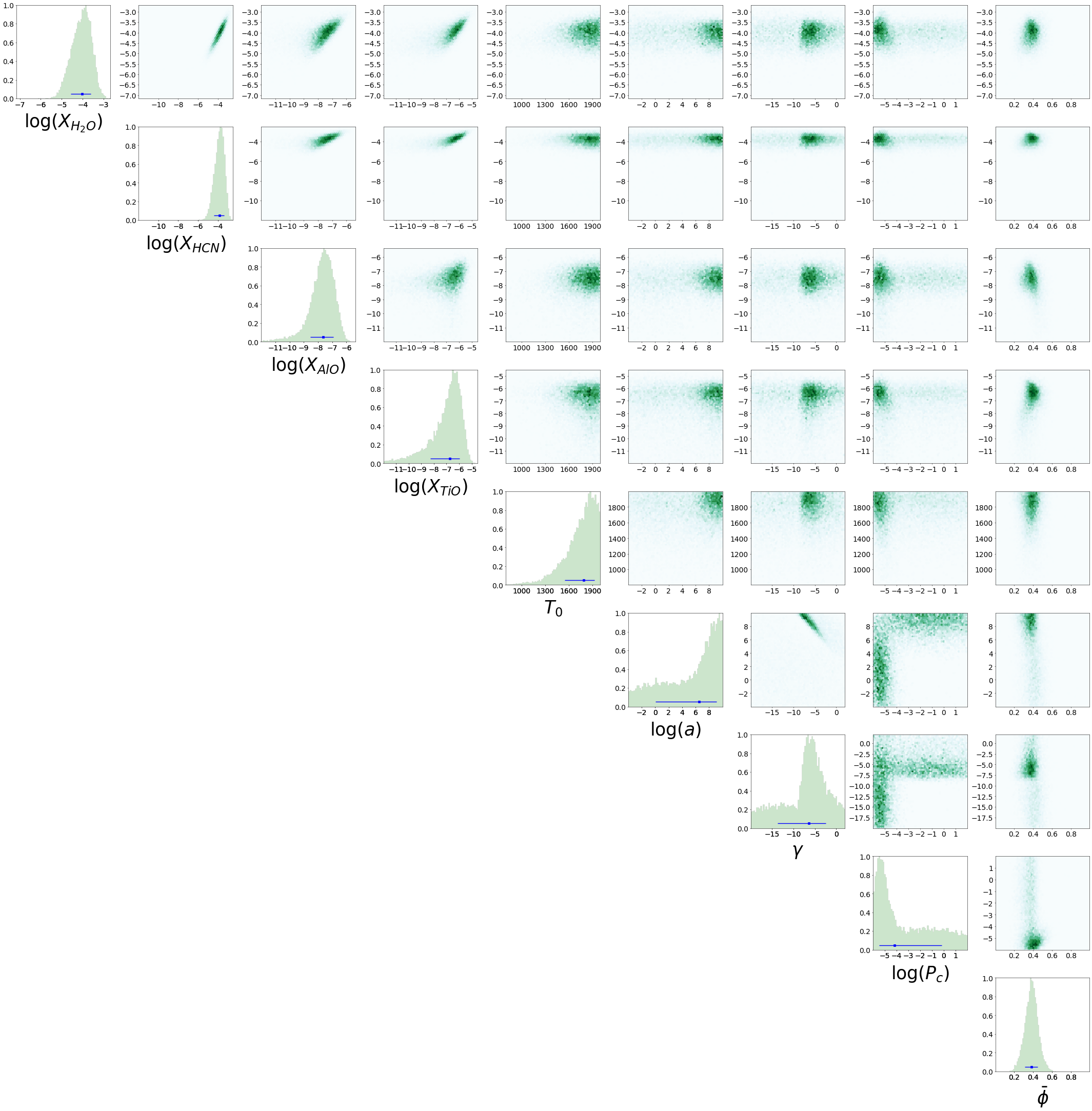}
\caption{Posterior distributions for the constrained chemical species in the fiducial model retrieval of KELT-11b. Temperature at the top of the atmosphere T$_0$ from the P-T profile and cloud/hazes parameters are also shown.
\label{fig:aura_posteriors_full}}
\end{center}
\end{figure*}

The retrieved model and observations are shown in Figure \ref{fig:aura_full}. The posterior distributions for the constrained chemical species, temperature at the top of the atmosphere (T$_0$), and cloud parameters are shown in Figure \ref{fig:aura_posteriors_full}. The retrieval finds a strong detection of H$_2$O at 3.6$\sigma$ with an abundance of $\log_{10}$(X$_{\text{H}_2{\text{O}}}$)$=-4.03 ^{+ 0.43 }_{- 0.53 }$ and indications of HCN at 2.7$\sigma$ with an abundance of $\log_{10}$(X$_{\text{HCN}}$)$=-3.84 ^{+ 0.45 }_{- 0.56 }$ based on the \HST/WFC3 transmission spectrum. The bluest part of the transmission spectrum and higher transit depth of the \TESS~data point relative to the \HST/WFC3 observations are preferentially explained by AlO or TiO, at 2$\sigma$ and 0.9$\sigma$ respectively. The retrieved abundances are $\log_{10}$(X$_{\text{AlO}}$)$=-7.64 ^{+ 0.71 }_{- 0.90 }$ and $\log_{10}$(X$_{\text{TiO}}$)$=-6.75 ^{+ 0.78 }_{- 1.53 }$.

Our fiducial model does not find strong constraints on the P-T profile of the atmosphere of KELT-11b or the presence of clouds and hazes.  We retrieve a temperature near the photosphere at 100mbar of T=$1982^{+341}_{-184}$K. Replacing the parametric P-T profile for an isothermal profile in our fiducial model results in a decrease in model evidence equivalent to 1.5$\sigma$. Similarly, removing inhomogenous clouds and hazes from our model in favor of a clear atmosphere results in a decrease of the model evidence equivalent to a 1.8$\sigma$ level. Neither decrease in model evidence is significant enough to robustly claim constraints on the P-T profile or the presence of clouds and hazes nor can they be confidently ruled out. 

To further consider the robustness of these inferences, we consider the possibility of an error-bar inflation free parameter  \citep[e.g.,][]{foremanmackey13}. Using this approach, it is assumed that the variance is underestimated by some fractional amount $\mathrm{f}$, namely

\begin{equation}
    S^2=\sigma_{\mathrm{obs}}^2+\mathrm{f}^2 \, \Delta_{\mathrm{mod}}^2
\end{equation}

\noindent where $\sigma_{\mathrm{obs}}$ is the error in the observations and $\Delta_{\mathrm{mod}}$ is the model's transit depth. Following this approach results in an increase in model evidence relative to our fiducial model equivalent to a 1.6$\sigma$ level. This means that the additional parameter is preferred at 1.6$\sigma$. The best fit model results in a $\chi^{2}$ of  23.40  for  5 degrees of freedom compared to the $\chi^{2}$ of  29.97  for  6 degrees of freedom in the fiducial model. The p-value is 2.83$\times10^{-04}$ and the BIC is 79.84, while for the fiducial model the p-value is 3.98$\times10^{-05}$ and the BIC is  83.27. The retrieved value for the error-inflation factor is $\log_{10}(\mathrm{f})=-1.93 ^{+ 0.12 }_{- 0.11 }$. While considering the possibility of an error-bar inflation factor results in better fits to the data, the inferred H$_2$O and HCN abundances are still consistent with those from the fiducial model. The retrieved abundances are $\log_{10}$(X$_{\text{H}_2{\text{O}}}$)$=-4.49 ^{+ 0.63 }_{- 0.84 }$ and $\log_{10}$(X$_{\text{HCN}}$)$=-4.62 ^{+ 0.94 }_{- 3.95 }$. When considering the error-bar inflation factor, H$_2$O and HCN are still preferred by the model at a 3.1$\sigma$ detection and 1.7$\sigma$ inference respectively. On the other hand, AlO and TiO are not preferred by the model. In the error-bar inflation model, the higher transit depth of the \TESS~data point can be explained by any of the species with signatures in the optical, namely Na, K, TiO, AlO, with no species being strongly preferred over other. 

We perform an additional set of retrievals on the complete transmission spectrum of KELT-11b considering the possibility of instrumental vertical offsets. We include two additional free parameters corresponding to possible offsets in transit depth in the \TESS~optical and \HST/WFC3 near-infrared bands relative to the \Spitzer~infrared bands. We consider two treatments for the prior on the offsets -- uniform and Gaussian priors. One set of retrievals considers a uniform prior on each offset ranging between [-80,80] ppm. Another set of retrievals assume the prior distribution to be a Gaussian centered on zero with a standard deviation ($\sigma$) of 80 ppm. A possible unaccounted shift of 80 ppm or higher is both generous and unlikely considering 80 ppm is $\sim$1.5$\times$ the precision of the \TESS~observations and $\gtrsim$3$\times$ the average precision of the \HST~observations. When considering these possible offsets, our results remain mostly unchanged with molecular abundances consistent with those obtained using the fiducial model. The presence of vertical offsets results in slightly better constraints on the abundance of TiO. On the other hand, the retrieved H$_2$O abundances remain unchanged. These results suggest that the retrieved molecular abundances using the fiducial model are robust against possible instrumental offsets and that our reported TiO abundance is a conservative estimate.

\subsubsection{Analysis of the \HST/WFC3~Transmission Spectrum}

\begin{figure}[h!]
\begin{center}
\includegraphics[width=0.5\textwidth]{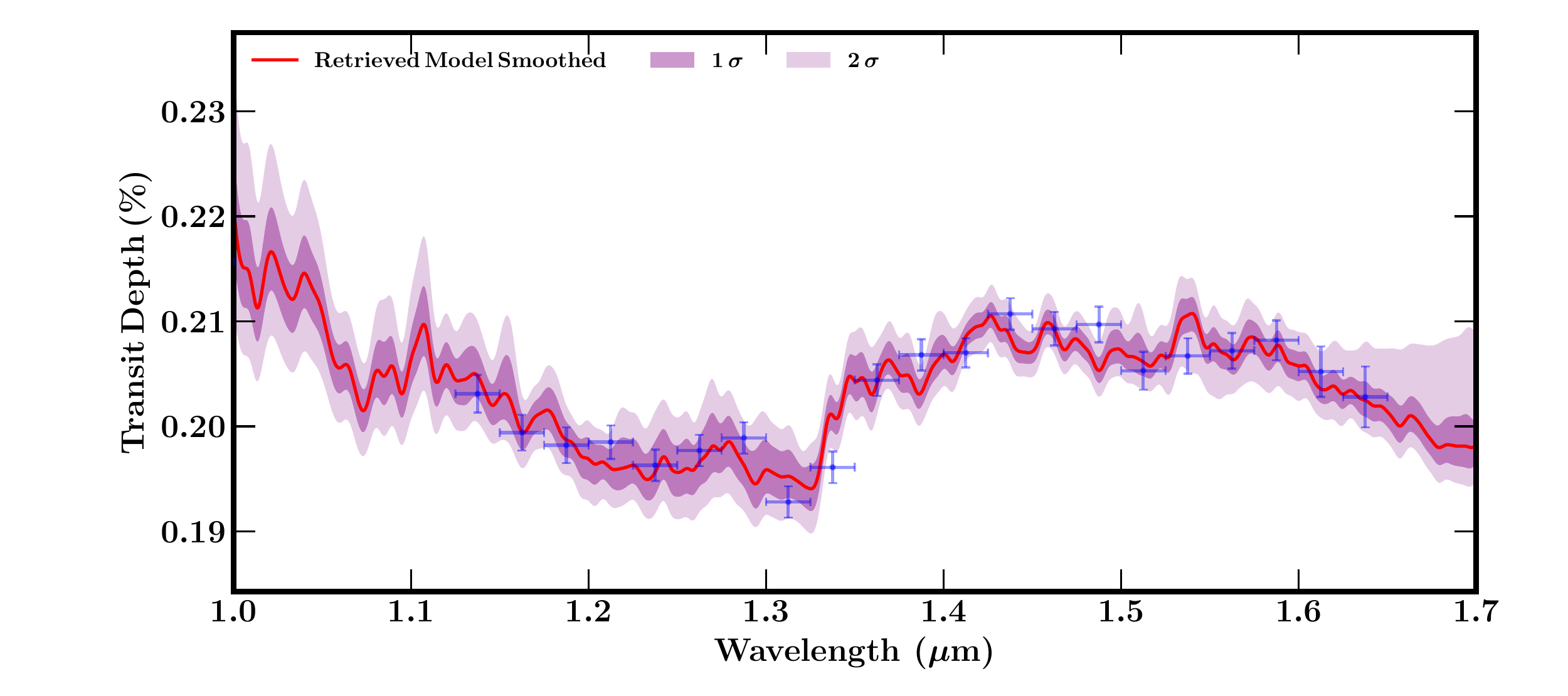}
\caption{Retrieved transmission spectrum of KELT-11b for the 17 parameter fiducial model using \HST/WFC3 observations only. The retrieved median transmission spectrum is shown in red with 1$\sigma$ and 2$\sigma$ contours shown in purple shaded regions. \HST/WFC3 observations are shown using blue markers. The best fit model has a $\chi^{2}$ of  23.28  for  4 degrees of freedom. The p-value is  1.11$\times10^{-04}$ and the BIC is 75.03.
\label{fig:aura_wfc3_full}}
\end{center}
\end{figure}

\begin{figure}[h!]
\begin{center}
\includegraphics[width=0.5\textwidth]{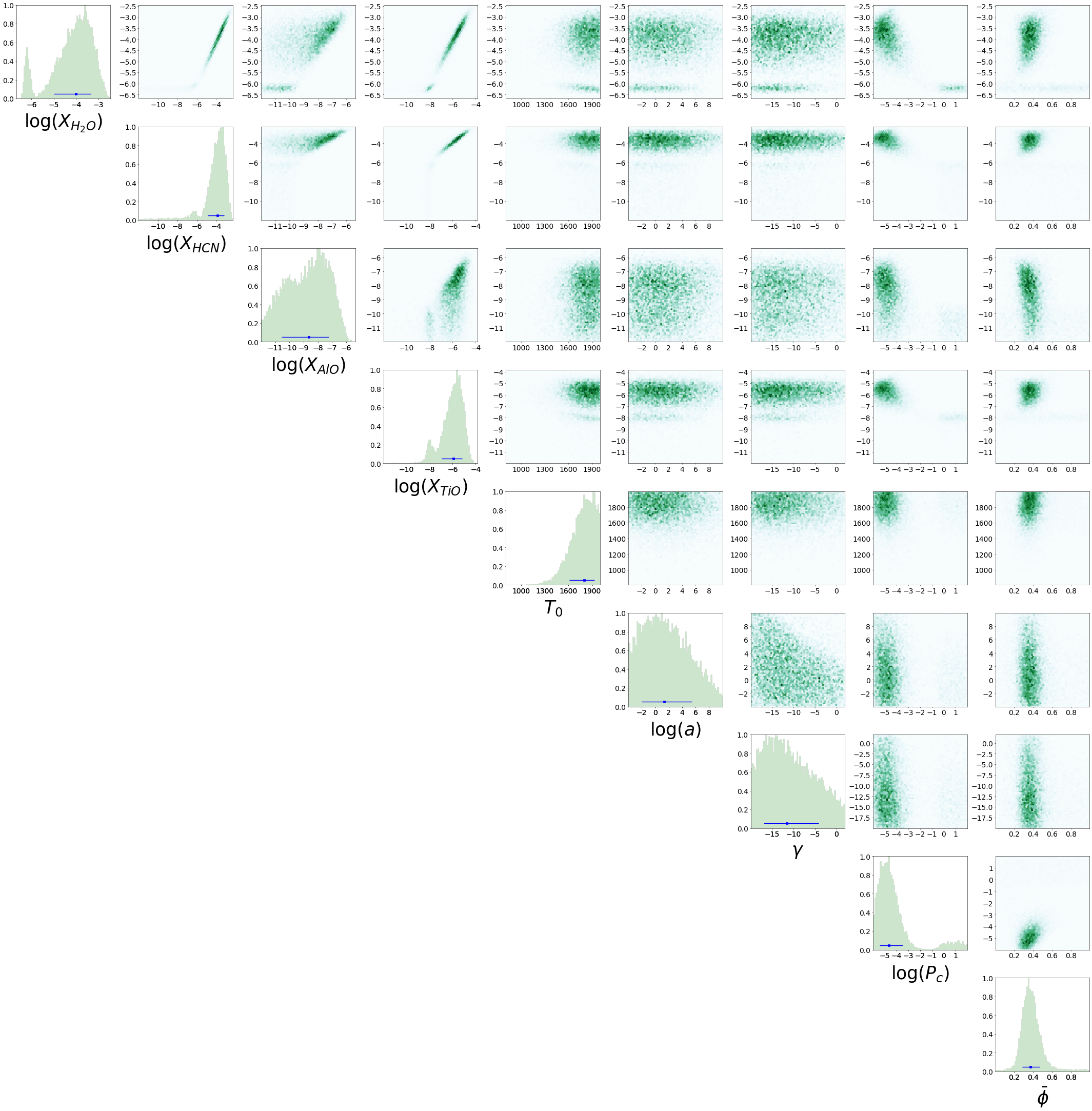}
\caption{Posterior distributions for the retrieval of KELT-11b using \HST/WFC3 observations only. The constrained chemical species are shown. Posterior distributions for the temperature at the top of the atmosphere T$_0$ and cloud/hazes parameters are included.
\label{fig:aura_posteriors_wfc3_full}}
\end{center}
\end{figure}

We further investigate the inferred chemical abundances and detections in KELT-11b when considering the \HST/WFC3 observations alone. We perform a retrieval on the \HST/WFC3 transmission spectrum using the same 17 parameter fiducial model described above. The retrieved transmission spectrum is shown in Figure \ref{fig:aura_wfc3_full}. This retrieval confirms the strong detection of H$_2$O at a confidence level of 4.6$\sigma$ with an abundance of $\log_{10}$(X$_{\text{H}_2{\text{O}}}$)$=-4.01 ^{+ 0.67 }_{- 0.98 }$. As in the retrieval of the complete transmission spectrum, this retrieval also explains the \HST/WFC3 observations with HCN absorption at a preference level of 2.5$\sigma$ and with an abundance of $\log_{10}$(X$_{\text{HCN}}$)$=-3.84 ^{+ 0.69 }_{- 1.01 }$. In contrast to the retrieval using the complete set of observations, this retrieval prefers TiO over AlO to explain the bluest spectral points in the \HST/WFC3 observations. The retrieved TiO abundance is $\log_{10}$(X$_{\text{TiO}}$)$=-5.91 ^{+ 0.73 }_{- 1.06 }$ at a detection significance of 2.9$\sigma$. On the other hand, the retrieved AlO abundance is $\log_{10}$(X$_{\text{AlO}}$)$=-8.65 ^{+ 1.43 }_{- 1.91 }$. When using the \HST/WFC3 observations only, removing AlO from the model results in an increase in model evidence indicating absorption due to this species is not preferred by the data. The cloud/hazes parameters and the temperature profile remain mostly unconstrained, with a retrieved temperature at 100mbar of T=$1959^{+242}_{-157}$K. The posterior distributions for the relevant parameters are shown in Figure \ref{fig:aura_posteriors_wfc3_full}.

Similarly to our analysis of the full transmission spectrum, we consider the possibility of underestimated error bars by retrieving an error-bar inflation parameter on the \HST/WFC3 observations only. This approach retrieves an error-bar inflation factor of $\log_{10}(\mathrm{f})=-2.07 ^{+ 0.14 }_{- 0.16 }$ and abundances of $\log_{10}$(X$_{\text{H}_2{\text{O}}}$)$=-4.73 ^{+ 1.13 }_{- 1.51 }$, $\log_{10}$(X$_{\text{HCN}}$)$=-4.52 ^{+ 1.15 }_{- 4.17 }$, $\log_{10}$(X$_{\text{TiO}}$)$=-7.24 ^{+ 1.75 }_{- 1.93 }$, and $\log_{10}$(X$_{\text{AlO}}$)$=-9.86 ^{+ 1.93 }_{- 1.27 }$, consistent with the non-inflated error-bar approach. Considering the error-bar inflation factor results in a decrease in the model evidence relative to the fiducial model without error-bar inflation comparable to a 2.8$\sigma$ level. This means that the additional error-bar inflation is not preferred at a 2.8$\sigma$ level. Similarly, this approach does not result in a better fit to the data by some frequentist metrics. The best fit model goes from a $\chi^{2}$ of  23.28  for 4 degrees of freedom in the non-inflated error-bar approach to a $\chi^{2}$ of  20.42  for  3 degrees of freedom when considering error-bar inflation. The p-value and BIC go from 1.11$\times10^{-04}$ and 75.03 respectively in the non-inflated error-bar approach to 1.39$\times10^{-04}$ and 75.22 when fitting for an error-bar inflation parameter.

\begin{figure}[h!]
\begin{center}
\includegraphics[width=0.5\textwidth]{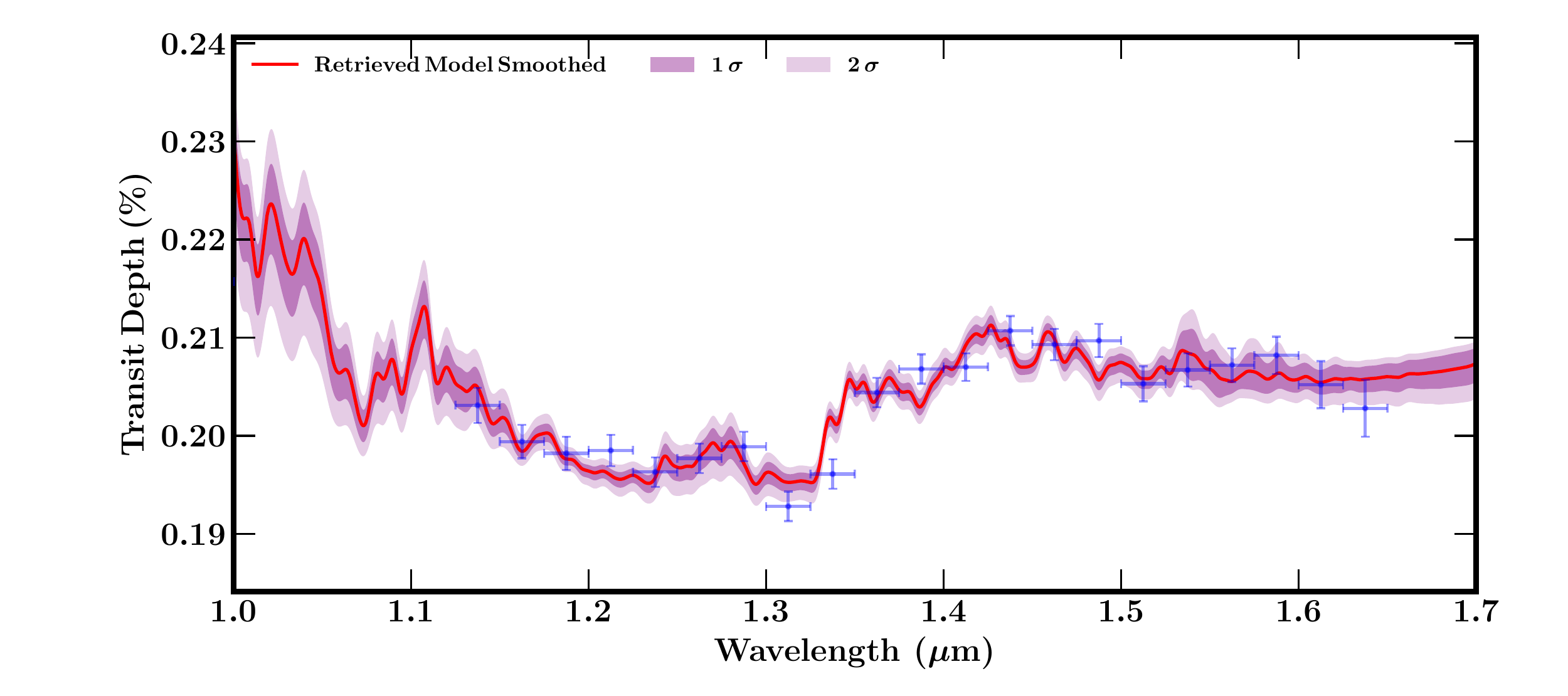}
\caption{Retrieved transmission spectrum of KELT-11b for an isothermal and clear atmosphere with limited absorbers using \HST/WFC3 observations only. The retrieved median transmission spectrum is shown in red with 1$\sigma$ and 2$\sigma$ contours shown in purple shaded regions. \HST/WFC3 observations are shown using blue markers. The best fit model has a $\chi^{2}$ of  28.85  for  16 degrees of freedom. The p-value is  2.50$\times10^{-02}$ and the BIC is  44.07.
\label{fig:aura_wfc3_simple}}
\end{center}
\end{figure}

\begin{figure}[h!]
\begin{center}
\includegraphics[width=0.5\textwidth]{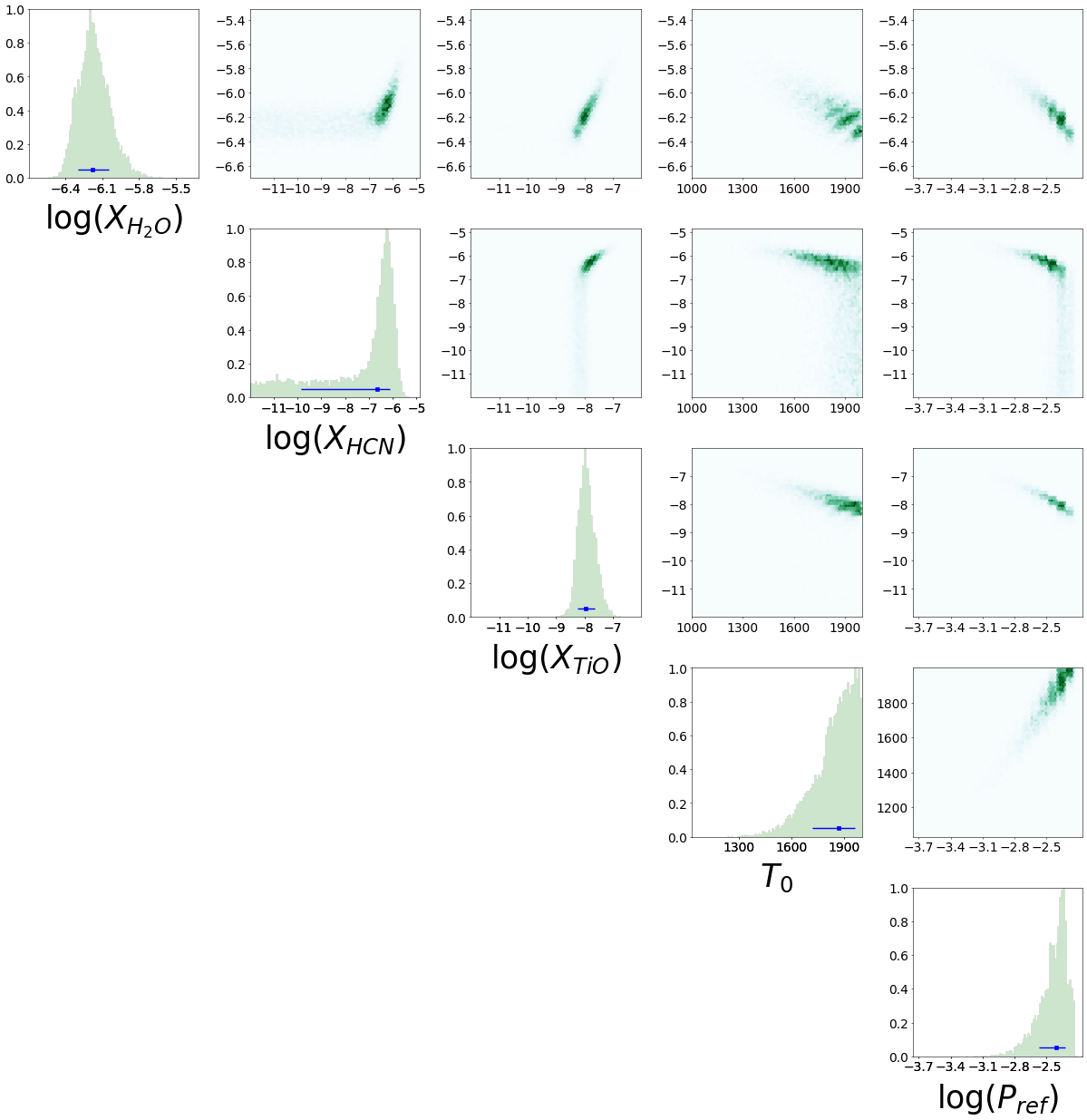}
\caption{Posterior distributions for the retrieval of KELT-11b using \HST/WFC3 observations only and assuming a clear and isothermal atmosphere.
\label{fig:aura_posteriors_wfc3_simple}}
\end{center}
\end{figure}

Lastly, given that current observations do not place strong constraints on the P-T profile or cloud and haze cover in the atmosphere of KELT-11b, we investigate the effects of considering a simpler model in the retrieved abundance estimates using the \HST/WFC3 observations only. We retrieve the atmospheric properties of KELT-11b using an isothermal P-T profile, a clear atmosphere, and considering absorption due to H$_2$O, HCN, and TiO. This retrieval results in abundances of $\log_{10}$(X$_{\text{H}_2{\text{O}}}$)$=-6.18 ^{+ 0.13 }_{- 0.12 }$, $\log_{10}$(X$_{\text{HCN}}$)$=-6.65 ^{+ 0.53 }_{- 3.20 }$, and $\log_{10}$(X$_{\text{TiO}}$)$=-7.96 ^{+ 0.33 }_{- 0.27 }$. While the median H$_2$O abundance is lower than that retrieved under different model considerations, the retrieved value is consistent with the estimates above. The abundances of HCN and TiO are also consistent with the estimates from the different model configurations. The TiO abundance is tightly over-constrained under these simplified considerations. The retrieved isothermal temperature is consistent with previous estimates with a value of T=$1867 ^{+ 93 }_{- 152 }$K. The retrieved transmission spectrum and posterior distributions are shown in Figures \ref{fig:aura_wfc3_simple} and \ref{fig:aura_posteriors_wfc3_simple} respectively. The reduced number of parameters results in a larger number of degrees of freedom which in turn translates to a better fit by frequentist metrics. The best fit model has a $\chi^{2}$ of  28.85  for  16 degrees of freedom. The p-value is  2.50$\times10^{-02}$ and the BIC is 44.07.

\begin{figure}[h!]
\begin{center}
\includegraphics[width=0.5\textwidth]{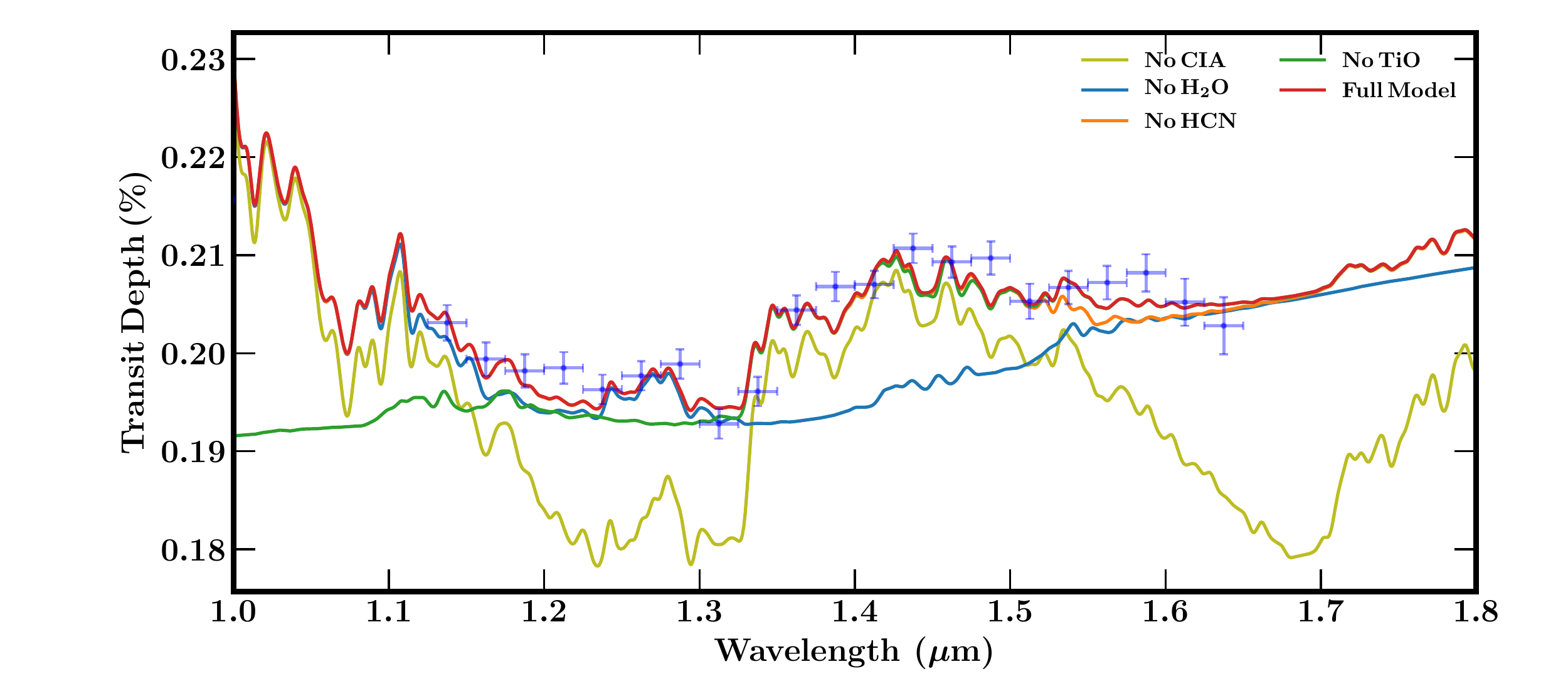}
\caption{Contribution plot for the simplified, 5 parameter model on the \HST/WFC3 observations. The red line shows a forward model with the median retrieved parameters. The blue, orange, green, and olive lines line shows the model without the contribution of H$_2$O, HCN, TiO, and H$_2$-H$_2$ and H$_2$-He CIA. \HST/WFC3 observations are shown using blue markers. 
\label{fig:aura_contribution}}
\end{center}
\end{figure}

We use these retrieved model parameters to produce a contribution plot. Figure \ref{fig:aura_contribution} shows the contribution of H$_2$O, HCN, and TiO to explain the \HST/WFC3 observations using the  retrieved median values from the simplified model above. The red curve in Figure \ref{fig:aura_contribution} shows the contribution to the transmission spectrum of all three species. The blue, orange, and green curves show the model without the contribution of H$_2$O, HCN, and TiO respectively. In olive, we show the contribution due to H$_2$-H$_2$ and H$_2$-He CIA.  From this figure, it can be seen that the H$_2$O contribution to the model is used to fit the spectral feature at $\sim$1.4 $\mu$m. On the other hand, the red-most part of the transmission spectrum at $\gtrsim$1.5$\mu$m is unusually flat (compared to typical transmission spectra; e.g., \citealt{iyer2016,tsiaras2018}) and is explained by HCN and CIA. The blue-most part of the transmission spectrum is being fit by absorption due to TiO.

For completion, we run one more retrieval considering the error-bar inflation parameter on the simplified model using the \HST/WFC3 observations only. The retrieved abundances are consistent with the non-inflated error-bar model. These are $\log_{10}$(X$_{\text{H}_2{\text{O}}}$)$=-6.18 ^{+ 0.14 }_{- 0.12 }$, $\log_{10}$(X$_{\text{HCN}}$)$=-6.79 ^{+ 0.62 }_{- 3.04 }$, and $\log_{10}$(X$_{\text{TiO}}$)$=-8.01 ^{+ 0.31 }_{- 0.33 }$. The retrieved error-bar inflation factor is $\log_{10}(\mathrm{f})=-3.41 ^{+ 1.35 }_{- 4.17 }$. Following this approach results in an even better fit by frequentist metrics. The best fit model has a $\chi^{2}$ of  18.66  for  15 degrees of freedom. The p-value is  2.30e-01 and the BIC is 36.92. The model evidence for the model with error-bar inflation is slightly smaller, $\sim$1$\sigma$ level, than that of the model without error-bar inflation. This decrease in model evidence indicates that the use of the additional error-bar inflation parameter is not preferred from a Bayesian perspective. We note that in this particular case while the inclusion of error inflation results in better fits to the data, i.e., better $\chi^2$ due to larger uncertainties on the data, the precision on the retrieved abundance is not highly affected. This is due to the use of simplified isothermal and cloud-free models on WFC3 data only that result in lower abundance estimates with small uncertainties \citep[see e.g.,][]{Welbanks2019a}.

\subsection{Transmission Spectrum Analysis with CHIMERA}\label{chimera}
Here we use the CHIMERA transmission retrieval tool \citep{Line2013, Line2016, kreidberg15, kreidberg2018} to explore a gradient in ``self-consistent" assumptions.  These more self-consistent methods are entirely complementary to the free retrieval analysis above.  The first approach is the ``chemically-consistent (CC)" method whereby chemical-equilibrium is assumed along a flexible temperature-pressure profile (one-way self-consistent). The second is a fit based on a small grid of self-consistent 1D-radiative convective models (two-way self-consistent).

\subsubsection{Chemically-Consistent Method}
In the CC framework we assume thermochemical equilibrium (no rainout) mixing ratios computed with the NASA CEA2 routine \citep{Gordon1994} along a T-P profile. For the T-P profile we assume a three-parameter version of the \citet{ParmentierGuillot14} analytic T-P profile framework, as implemented in \cite{Line2013}, given the metallicity ([M/H]) and carbon-to-oxygen ratio (C/O). Abundances are scaled relative to solar composition based on  \cite{Lodders2003}. Specifically, all elements are first re-normalized relative to H to preserve the solar abundance pattern, and then the C/O is adjusted preserving the sum of C and O at the scaled metallicity value.  Equilibrium composition is computed for hundreds of molecules, atoms, condensates, and ions; however we only include the opacity from H$_2$O \citep{Partridge1997}, CH$_4$ \citep{Yurchenko2014}, CO, CO$_2$ \citep{Rothman2010}, NH$_3$ \citep{Yurchenko2011},   VO \citep{McKemmish2016}, TiO \citep{McKemmish2019}, C$_2$H$_2$, HCN \citep{Harris2006}, H$_2$S, FeH \citep{Hargreaves2010}, PH$_3$, SiO \citep{Barton2013}, H$_2$-H$_2$/He CIA \citep{Richard2012} and H$_2$/He molecular Rayleigh scattering.

We also test two different cloud parameterizations: the first is the ``classic" power-law haze, with haze amplitude and slope as in Section \ref{aura} above, plus a cloud with a single vertically uniform gray opacity. In total there are three free parameters. The second model is the \cite{Ackerman2001} pseudo-microphysical cloud framework (assuming enstatite grains) as described and implemented in \cite{MaiLine19} (constant eddy diffusion coefficient, cloud sedimentation parameter, cloud base pressure, and cloud base condensate mixing ratio). In all cases we also retrieve for the 10 bar radius.  

In total, the free parameters include the three controlling the T-P profile, the composition parameters [M/H] and C/O, the 10 bar radius, and three or four cloud parameters (9-10 parameters total).  The  uniform (or log-uniform) prior ranges are generous. The tightest restriction is on the irradiation temperature, which is specified not to exceed much more than the planetary equilibrium temperature (up to 1800~K). All parameter estimates are determined with the {\tt PyMultiNest} tool \citep{Buchner2014}.     

\begin{figure*}
\begin{center}
\includegraphics[width=1.0\textwidth]{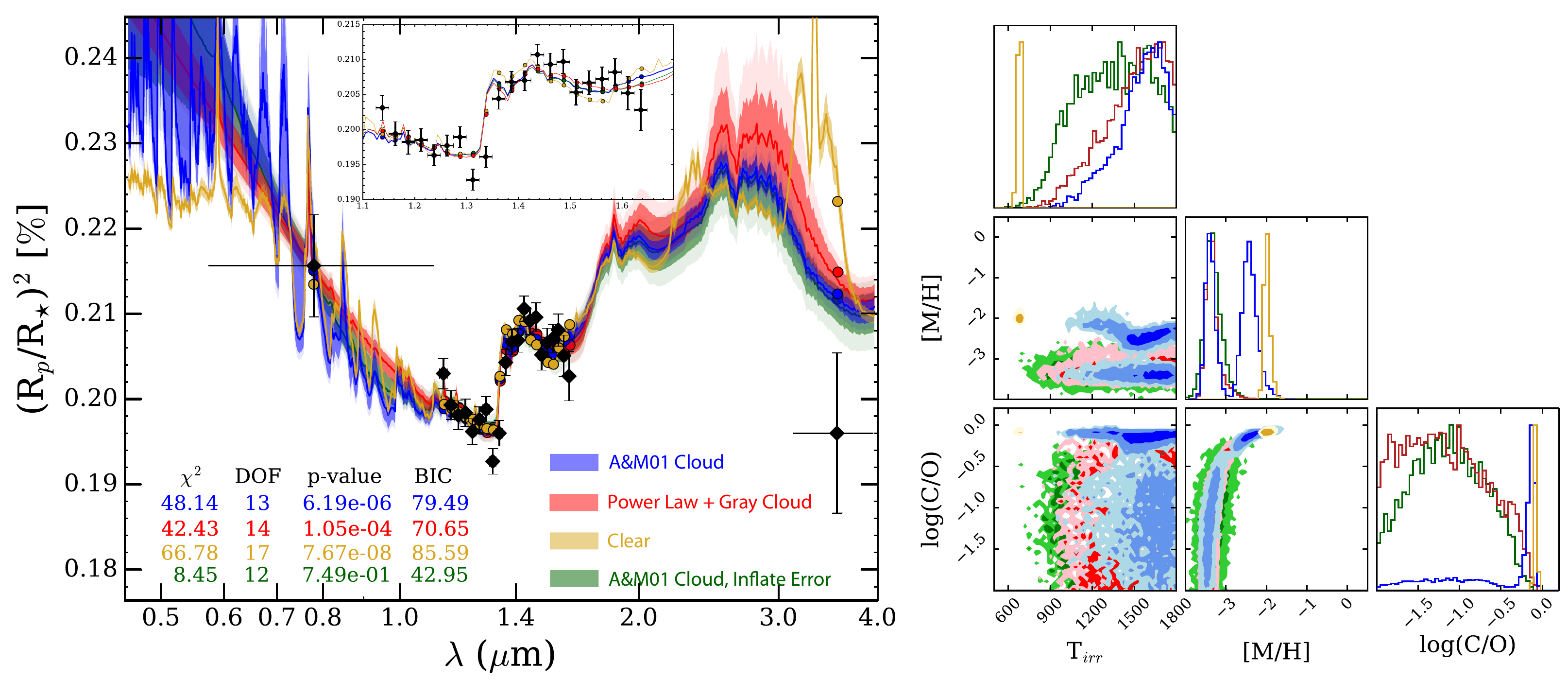}
\caption{Summary of the chemically-consistent retrieval analysis under four different scenarios (blue: \cite{Ackerman2001} cloud parameterization, red: power law + gray cloud, orange: clear, green: \cite{Ackerman2001} cloud parameterization but with a \HST/WFC3 error bar inflation term). The fits (left) are summarized with a median (and binned median, points), and 1$\sigma$ and 2$\sigma$ credibility regions from 250 posterior draws. The inset shows a zoom in of the \HST/WFC3 data (with median fits only).  The right panel summarizes the TP profile irradiation temperature, the metallicity, and C-to-O under the different scenarios.  They all lead to extreme low metallicities, and an inconsequential upper limit on the C/O (with the exception of the clear scenario).  However, all of the fits are rather poor by any standard frequentist means, with even the best fits being strongly rejected by the data.  The error-bar inflation scenario retrieves an inflation scale factor to the \HST/WFC3 data of 1.7 ($\pm0.28$), either due to missing model aspects, or underestimated error bars.       }
\label{fig:chimera_cc_alldata}
\end{center}
\end{figure*}

Within the CC setup we explore the following four scenarios and their influence on the retrieved [M/H] and C/O from the combined \TESS+\HST+\Spitzer~transmission spectrum: 
\begin{itemize}
    \item A\&M01 cloud: the nominal 10-parameter model with the \citet{Ackerman2001} parameterization
    \item Power law plus gray cloud: same as the nominal case but instead of the \citet{Ackerman2001} parameterization, the three parameter power-law-haze and gray cloud parameterization is instead used (eight parameters)
    \item Clear: Same as nominal, except the cloud parameters are fixed to produce no opacity (six parameters)
    \item A\&M01 cloud with inflated error bars: the 10-parameter model with the \citet{Ackerman2001} parameterization, but with an additional error bar inflation term to the \HST/WFC3 data (implemented as a constant scaling factor free parameter to the data-errors and the accordingly modified log-likelihood)
\end{itemize}
Figure \ref{fig:chimera_cc_alldata} summarizes the fits and relevant constraints under these four scenarios. We focus on the composition constraints as the other parameters are largely uninformative and are considered nuisance parameters.  The most striking find from this analysis is the extremely low metallicity, $[M/H] \lesssim -2$, in all cases.  This constraint is primarily driven by the unusually ``flat" spectral shape between 1.5 and 1.65 $\mu$m in the \HST/WFC3 data.  In most transmission spectra \citep[e.g.,][]{iyer2016,tsiaras2018} (provided they are not completely flat), the transit depths tends to decrease around 1.5--1.65 $\mu$m, and when combined with the larger depths near 1.32~$\mu$m these features indicate the presence of water.  The roughly constant transit depth redder than 1.4 $\mu$m in the \HST/WFC3 transmission spectrum of KELT-11b is suggestive of additional opacity.  In the free retrievals with AURA (Section \ref{aura}), that additional opacity is due to HCN along with H$_2$ collision induced opacity. However, given that HCN is not particularly abundant in thermochemical equilibrium, the CC retrieval attempts to find an alternate solution without HCN. Rather than add opacity, the CC retrieval seeks to remove H$_2$O opacity, essentially making the H$_2$ CIA slope more visible. Increased C/O, in contrast, would have added CH$_4$ and HCN opacity, which, together, were less favored by the data (thus the upper limit on the C/O we find here in the CC retrieval). The retrieved metallicity changes slightly depending on the cloud model assumption, but still remains low.  The cloud-free scenario results in very cold terminator temperatures and a highly constrained metallicity and C/O.  

It is important to note that all three scenarios produce poor fits, with low p-values ($<10^{-3}$). This implies that scenarios 1-3 are all overwhelmingly rejected by the data,  obviating the constraints on the atmospheric composition. To remedy this issue,  the fourth scenario (with a model identical to the first scenario), includes an error-bar inflation free parameter \citep[e.g.,][]{Line2015} to account for data-model mis-matches. This inflation scale factor is about 1.7 over the \HST/WFC3 bandpass, resulting in acceptable p-values. This inflation naturally leads to larger uncertainties on the parameter constraints, though the low metallicity solution still stands.

We also performed a similar analysis on the \HST/WFC3 data alone (not shown) and arrived at a similar low metallicity conclusion, again being driven by the shape of the red edge of the \HST/WFC3 transmission spectrum.

\subsubsection{Self-Consistent 1D Radiative-Convective Equilibrium Model Grid Fit}
\label{chimera:rc}

In addition to the CC retrievals, we also explore self-consistent 1D-radiative convective equilibrium fits (1D-RC).  We use the 1D-RC model, ScCHIMERA, described in \cite{pis18, Arcangeli2018, GharibLine19} to generate a KELT-11b-specific self-consistent TP-profile/chemistry (cloud-free) grid over a metallicity and irradiation temperature (a proxy for redistribution). The TP-profiles and chemistry from the converged models are then used to ``post-process" transmission spectra (assuming that that 1D solution represents the entire planet--a reasonable assumption given that the 4.5 $\mu$m dayside eclipse depth is suggestive of full redistribution--described below) while adding in the simple power law+gray cloud (and accounting for the 10 bar radius parameter) on the fly, with a nearest neighbor grid-search within the nested sampling. 

\begin{figure*}
\begin{center}
\includegraphics[width=1.0\textwidth]{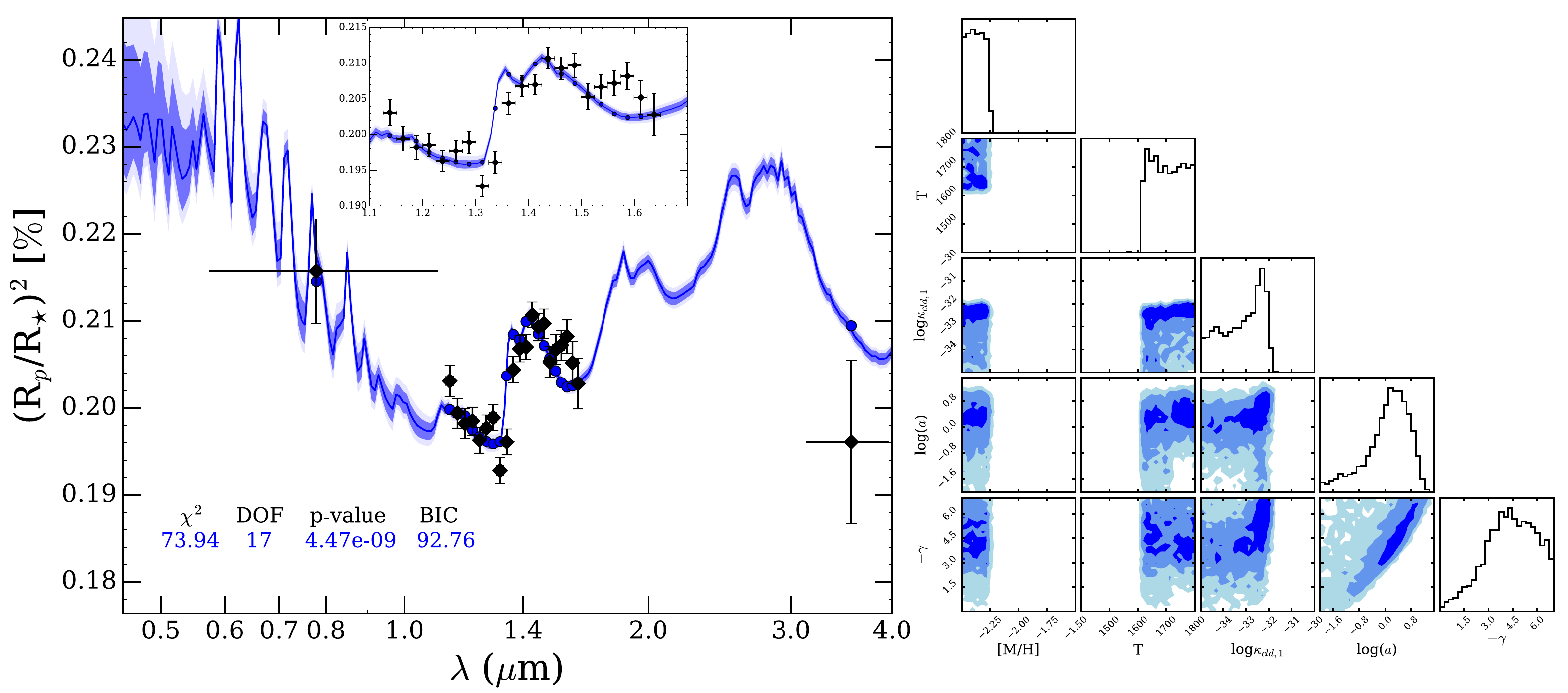}
\caption{Summary of the 1D self-consistent radiative-convective equilibrium grid ``post-processed" retrieval with the power law + gray cloud. The fits (left) are summarized with a median (and binned median, points), and 1$\sigma$ and 2$\sigma$ credibility regions from 250 posterior draws. The inset shows a zoom in of the \HST/WFC3 data.  The right panel summarizes the posterior (irradiation temperature: $T$, metallicity: $[M/H]$, gray cloud opacity: log$\kappa_{cld}$, haze amplitude: log($a$), and haze power law index: -$\gamma$). The primary conclusion, as with the CC retrievals, is the low metallicity, and more-or-less upper limits on cloud/haze opacities.  Given the poor fit to the data, the constraints on atmospheric properties should be treated with caution.}
\label{fig:self_cons_grid}
\end{center}
\end{figure*}

The results are summarized in Figure \ref{fig:self_cons_grid}. As with the CC retrievals, we find a notably low metallicity as well as primarily upper limits on the power law haze [log$(a)$] and gray cloud opacities (log$\kappa_{cld}$).  The fits are also poor, with the best being strongly rejected by the data.  We experimented with various assumptions within this 1D-RC ``post-processes" framework, that include error bar inflation (as above), east-west terminator variation (via an averaging of spectra with different temperatures--hence composition--and haze/cloud properties), as well as a re-generation of the 1D-RC grid, and fits, to account for quenching within the NH$_3$-N$_2$-HCN and CH$_4$-CO systems [assuming a constant log$K_{zz}$=10 (cgs) via \cite{ZahnleMarley2014}]. None of these substantially altered the resulting metallicity, and all of the fits (with the exception of the error-bar inflation, by construction) were equivalently poor, again suggesting constraints presented here should be interpreted with caution.   

\subsection{Summary of Results from AURA and CHIMERA for the Transmission Spectrum of KELT-11b}

\begin{table*}[]
\begin{footnotesize}
\begin{centering}
\begin{tabular}{cccccc}
Scenario & \begin{tabular}{@{}c@{}}Abundance Constraint \end{tabular} & D.O.F. & $\chi^2$ &  p-value & BIC \\
\hline
\hline
AURA (Free Retrievals): &  &  &  &  & \\
\hline
\hline
\shortstack{\strut Free 1\\(Full Model with Clouds/Hazes, All Data,\\Figs. \ref{fig:aura_full}, \ref{fig:aura_posteriors_full})} & \begin{tabular}{@{}c@{}}$\log_{10}$(X$_{\text{H}_2{\text{O}}}$)$=-4.03 ^{+ 0.43 }_{- 0.53 }$\\  $\log_{10}$(X$_{\text{HCN}}$)$=-3.84 ^{+ 0.45 }_{- 0.56 }$ \\ $\log_{10}$(X$_{\text{TiO}}$)$=-6.75 ^{+ 0.78 }_{- 1.53 }$ \\ $\log_{10}$(X$_{\text{AlO}}$)$=-7.64 ^{+ 0.71 }_{- 0.90 }$ \end{tabular} &  6  & 29.97 & 3.98$\times10^{-05}$ & 83.27 \\
\hline
\shortstack{\strut Free 2\\(Full Model with Clouds/Hazes, All Data, \\Error Inflation)} & \begin{tabular}{@{}c@{}}$\log_{10}$(X$_{\text{H}_2{\text{O}}}$)$=-4.49 ^{+ 0.63 }_{- 0.84 }$ \\ $\log_{10}$(X$_{\text{HCN}}$)$=-4.62 ^{+ 0.94 }_{- 3.95 }$  \end{tabular} & 5 & 23.40 & 2.83$\times10^{-04}$ & 79.84\\
\hline
\shortstack{\strut Free 3\\(Full Model with Clouds/Hazes, WFC3 Only,\\Figs. \ref{fig:aura_wfc3_full}, \ref{fig:aura_posteriors_wfc3_full})}& \begin{tabular}{@{}c@{}}$\log_{10}$(X$_{\text{H}_2{\text{O}}}$)$=-4.01 ^{+ 0.67 }_{- 0.98 }$\\ $\log_{10}$(X$_{\text{HCN}}$)$=-3.84 ^{+ 0.69 }_{- 1.01 }$ \\ $\log_{10}$(X$_{\text{TiO}}$)$=-5.91 ^{+ 0.73 }_{- 1.06 }$\\ $\log_{10}$(X$_{\text{AlO}}$)$=-8.65 ^{+ 1.43 }_{- 1.91 }$ \end{tabular} & 4 & 23.28 & 1.11$\times10^{-04}$ & 75.03 \\
 \hline
\shortstack{\strut Free 4\\(Full Model with Clouds/Hazes, WFC3 Only,\\Error Inflation)}& \begin{tabular}{@{}c@{}}$\log_{10}$(X$_{\text{H}_2{\text{O}}}$)$=-4.73 ^{+ 1.13 }_{- 1.51 }$\\ $\log_{10}$(X$_{\text{HCN}}$)$=-4.52 ^{+ 1.15 }_{- 4.17 }$ \\ $\log_{10}$(X$_{\text{TiO}}$)$=-7.24 ^{+ 1.75 }_{- 1.93 }$\\$\log_{10}$(X$_{\text{AlO}}$)$=-9.86 ^{+ 1.93 }_{- 1.27 }$\end{tabular} & 3 & 20.42 & 1.39$\times10^{-04}$ & 75.22 \\
 \hline
\shortstack{\strut Free 5\\(Isothermal/Cloud-Free/Simple Model, WFC3 Only,\\Figs. \ref{fig:aura_wfc3_simple}, \ref{fig:aura_posteriors_wfc3_simple})}& \begin{tabular}{@{}c@{}}$\log_{10}$(X$_{\text{H}_2{\text{O}}}$)$=-6.18 ^{+ 0.13 }_{- 0.12 }$ \\ $\log_{10}$(X$_{\text{HCN}}$)$=-6.65 ^{+ 0.53 }_{- 3.20 }$ \\ $\log_{10}$(X$_{\text{TiO}}$)$=-7.96 ^{+ 0.33 }_{- 0.27 }$\end{tabular}  & 16 & 28.85 & 2.50$\times10^{-02}$ & 44.07  \\
\hline
\shortstack{\strut Free 6\\(Isothermal/Cloud-Free/Simple Model, WFC3 Only,\\Error Inflation)} & 
\begin{tabular}{@{}c@{}}$\log_{10}$(X$_{\text{H}_2{\text{O}}}$)$=-6.18 ^{+ 0.14 }_{- 0.12 }$\\ $\log_{10}$(X$_{\text{HCN}}$)$=-6.79 ^{+ 0.62 }_{- 3.04 }$\\ $\log_{10}$(X$_{\text{TiO}}$)$=-8.01 ^{+ 0.31 }_{- 0.33 }$\end{tabular}  & 15 & 18.66 & 2.30$\times10^{-01}$ & 36.92  \\
\hline
\shortstack{\strut Free 7\\(Full Model with Clouds/Hazes, All Data,\\Gaussian prior vertical offset)} & \begin{tabular}{@{}c@{}}$\log_{10}$(X$_{\text{H}_2{\text{O}}}$)$=-4.08 ^{+ 0.46 }_{- 0.53 }$\\  $\log_{10}$(X$_{\text{HCN}}$)$=-3.99 ^{+ 0.51 }_{- 0.81 }$ \\ $\log_{10}$(X$_{\text{TiO}}$)$=-6.07 ^{+ 0.54 }_{- 0.64 }$ \\ $\log_{10}$(X$_{\text{AlO}}$)$=-7.95 ^{+ 0.86 }_{- 1.42 }$ \\ Shift$_{\mathrm{TESS}}=67.98 ^{+54.68  }_{- 52.00 }$ ppm \\ Shift$_{\mathrm{HST}}=-110.83 ^{+48.58 }_{-46.61}$ ppm  \end{tabular} & 4 &  27.66  & 1.46$\times10^{-05}$ & 87.23\\
\hline
\shortstack{\strut Free 8\\(Full Model with Clouds/Hazes, All Data,\\uniform prior vertical offset)} & \begin{tabular}{@{}c@{}}$\log_{10}$(X$_{\text{H}_2{\text{O}}}$)$=-3.98 ^{+ 0.44 }_{- 0.54 }$\\  $\log_{10}$(X$_{\text{HCN}}$)$=-3.82 ^{+ 0.45 }_{- 0.58 }$ \\ $\log_{10}$(X$_{\text{TiO}}$)$=-6.17 ^{+ 0.56 }_{- 0.76 }$ \\ $\log_{10}$(X$_{\text{AlO}}$)$=-7.87 ^{+ 0.87 }_{- 1.51 }$\\ Shift$_{\mathrm{TESS}}=45.49 ^{+ 23.84  }_{-40.94}$ ppm \\ Shift$_{\mathrm{HST}}=-58.93 ^{+27.36  }_{- 15.02 }$ ppm  \end{tabular} & 4 &  31.66 & 2.24$\times10^{-06}$ & 91.23\\
\hline
\hline
CHIMERA (Self-Consistent Retrievals): &  &  &  &  & \\
\hline
\hline
\shortstack{\strut CC1\\(A\&M01 Cloud, Fig. \ref{fig:chimera_cc_alldata})} & \begin{tabular}{@{}c@{}}$[M/H]=-2.61^{+ 0.30 }_{- 0.82 }$\end{tabular} & 13 & 48.14 & 6.19$\times10^{-06}$ & 79.49 \\
\hline
\shortstack{\strut CC2\\(Power Law + Gray Cloud, Fig. \ref{fig:chimera_cc_alldata})} & \begin{tabular}{@{}c@{}}$[M/H]=-3.36^{+ 0.17 }_{-0.12 }$\end{tabular} & 14 & 42.43 & 1.05$\times10^{-04}$ & 70.65 \\
\hline
\shortstack{\strut CC3\\(Clear, Fig. \ref{fig:chimera_cc_alldata})} & \begin{tabular}{@{}c@{}}$[M/H]=-2.00^{+ 0.09 }_{- 0.05 }$\end{tabular} & 17 & 66.78 & 7.67$\times10^{-08}$ & 85.59 \\
\hline
\shortstack{\strut CC4\\(A\&M01 Cloud, Error Inflation, Fig. \ref{fig:chimera_cc_alldata})} & \begin{tabular}{@{}c@{}}$[M/H]=-3.33^{+ 0.22 }_{- 0.18 }$\end{tabular} & 12 & 8.45 & 7.49$\times10^{-01}$ & 42.95 \\
\hline
\shortstack{\strut 1D-RC\\(1D-Radiative Convective Equilibrium, Fig. \ref{fig:self_cons_grid})} & \begin{tabular}{@{}c@{}}$[M/H] < -2.25$\end{tabular} & 17 & 73.94 & 4.47$\times10^{-09}$ & 92.76 \\

\end{tabular}
\caption{Summary of the different retrievals performed here. There are 23 data-points in the complete transmission spectrum (\TESS+\HST+\Spitzer) and 21 in the \HST~transmission spectrum. The CHIMERA retrievals included here were all performed on the full data set. For reference, solar abundances at 12 mbar and 1730~K are: $\log_{10}$(X$_{\text{H}_2{\text{O}}}$) = -3.44, $\log_{10}$(X$_{\text{HCN}}$) = -9.62, $\log_{10}$(X$_{\text{TiO}}$) = -11.63, $\log_{10}$(X$_{\text{AlO}}$) $<$ -14. The (un-inflated) free retrievals with AURA tend to produce $\chi^2/N_{data}$ between 1.1--1.3 whereas the self/chemically-consistent models (un-inflated) fall between 1.8--3.2.  The nominal H$_2$O abundances from the AURA retrievals are all sub-solar ($\sim$0.002--0.3$\times$solar) while the nominal HCN, TiO, and AlO abundances are all far out of equilibrium. The precisions on [M/H] should be interpreted with caution for the CHIMERA models given the poor fits, though they all consistently favor extreme-sub-solar values.}
\label{tab:retrievals}
\end{centering}
\end{footnotesize}
\end{table*}

Table \ref{tab:retrievals} provides a summary of the multiple retrievals we ran here. We find that none of the more self-consistent models using the CHIMERA retrieval tool (either the CC or the full 1D-RC) adequately fit the entire \TESS+\HST+\Spitzer~transmission data set.  All of the best fits over the range in plausible assumptions (clouds, terminator in-homogeneity, quenching) are considered strongly rejected by the data by any standard frequentist metric. The free retrieval experiments with the AURA retrieval tool, on the other hand, provide much more adequate fits overall ($\chi^2$ per data point $\sim$1.1--1.3). However, in many cases these also result in very low p-values (with the exception of the simple model on WFC3--Free 5/6--scenarios) given the large numbers of free parameters (hence small degrees of freedom). Furthermore, in several scenarios the abundance constraints are at odds with solar abundance patterns (see Table \ref{tab:retrievals} caption) at expected planetary temperatures. For example, in the Free 1 scenario, HCN, AlO, and TiO are $\gtrapprox$160,000, 300,000, and 2,000 solar expectations, respectively, whereas water is slightly subsolar--0.08--0.7$\times$ solar. 

Nevertheless, the AURA model consistently detected H$_2$O at $>$3.1$\sigma$ and inferred the presence of HCN at $>$1.7$\sigma$ to explain the unusually flat shape of the red end of the \HST~transmission spectrum. The fiducial AURA retrieval on the combined \TESS+\HST+\Spitzer~transmission spectrum also inferred the presence of AlO at 2$\sigma$, however, with error bar inflation AlO was no longer preferred by the model.  TiO is preferred to explain the shape of the blue end of the \HST~transmission spectrum, when performing a retrieval with AURA on the \HST~transmission spectrum alone. On the other hand, TiO is not preferred in the AURA retrieval of the combined \TESS+\HST+\Spitzer~transmission spectrum, but the larger transit depth measured from \TESS~compared to \HST~is generally suggestive of the presence of an additional optical absorber.

\subsection{Emission Spectrum Analysis with CHIMERA}\label{emission}
\begin{figure}[ht]
\begin{center}
\includegraphics[width=0.45\textwidth]{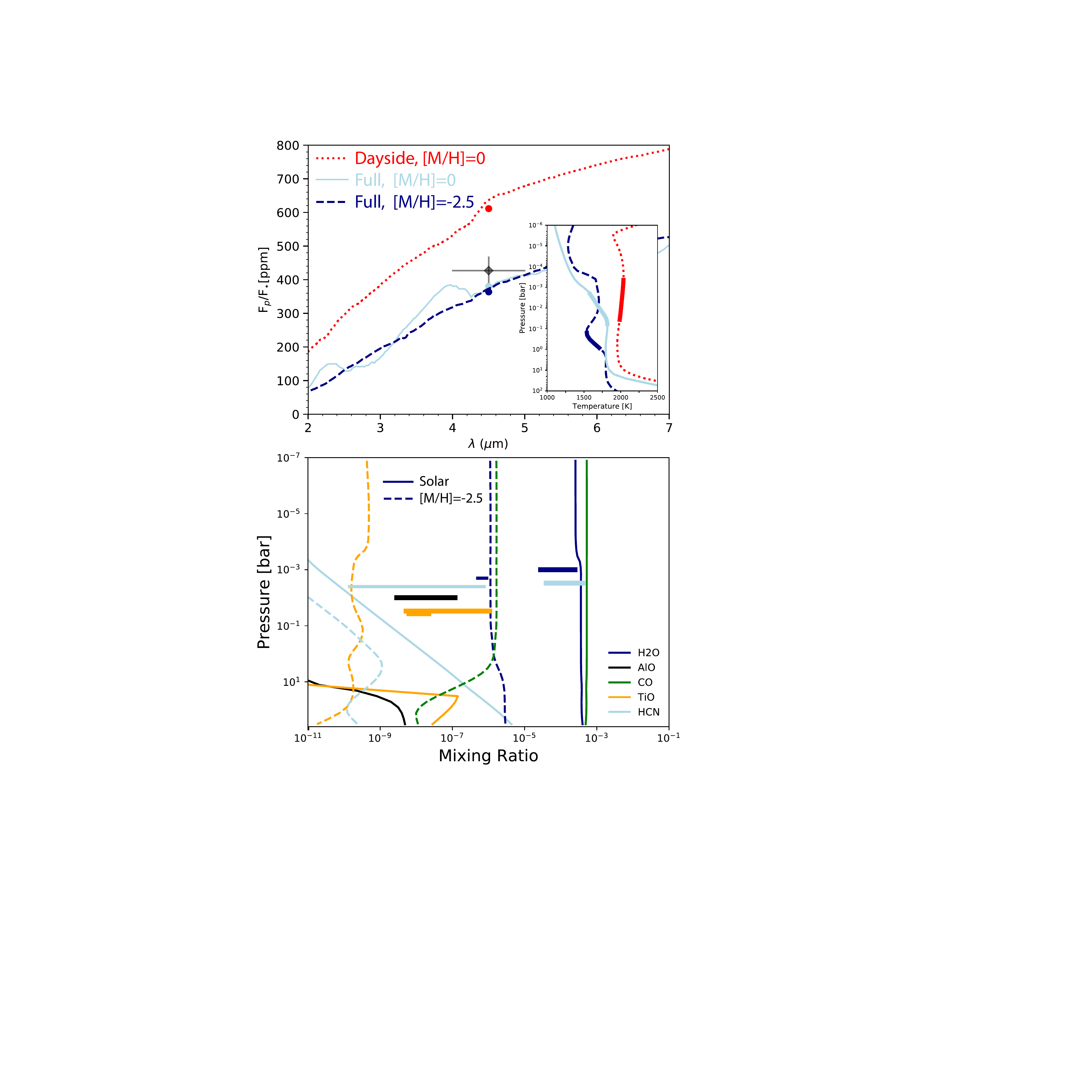}
\caption{Select 1D self-consistent atmospheric scenarios compared to the 4.5 $\mu$m eclipse depth (diamond w/ error bars).  The top panel compares the spectra (and TP profiles, inset) under a dayside redistribution solar scenario (red-dotted), full redistribution, solar scenario (light-blue, solid), and a full redistribution, low metallicity scenario (navy, dashed).  In the TP-profile inset, the 4.5 $\mu$m contribution functions are shown as solid line-segments along the profiles. The measured eclipse depth is more consistent with the full redistribution scenario regardless of the atmospheric metallicity.  The bottom panel compares the chemical-equilibrium mixing ratios under the full solar (solid) and low metallicity (dashed) scenarios.   The horizontal lines represent the 1$\sigma$ abundance constraints from the free retrievals (Free 1 as thick, Free 5 as thin) in Section \ref{aura} (with the location in pressure arbitrary as abundances are assumed constant in the atmosphere).       
\label{fig:emission}}
\end{center}
\end{figure}

We use the same set of 1D-RC grid models described above in Section \ref{chimera:rc} to provide insight into the 4.5 $\mu$m eclipse depth that we measured (427$\pm$42 ppm).  We do not run a full ``grid retrieval" on this single data point, but rather focus on a few representative scenarios: dayside redistribution (dayside temperature of $\sim$2000K), full redistribution (dayside temperature of $\sim$1700K), plus the latter with ultra-low metallicity (Figure \ref{fig:emission}). We note the 4.5 $\mu$m band sits within the well-known CO feature in a solar composition atmosphere (which can be seen subtly in emission in the dayside redistribution scenario model in Figure \ref{fig:emission}). The low metallicity eclipse model spectrum is consistent with a near pure blackbody as there is little opacity beyond H$_2$ collision induced continuum.  

We find that the 4.5 $\mu$m eclipse depth is more consistent with a full-distribution-like scenario regardless of metallicity (Figure \ref{fig:emission}, top).  This suggests that there could be little day-to-night temperature in-homogeneity, hence fairly uniform terminator temperatures/properties. However, we cannot completely rule out the dayside redistribution scenario because the dayside of the planet could have clouds, which would lower the Bond albedo and lead to a cooler dayside (and a smaller eclipse depth consistent with the full redistribution scenario). Eclipse measurements at additional wavelengths are needed to support our interpretation. We note that while the \TESS~observations presented in Figure \ref{fig:tess} do cover multiple eclipse windows, we can only place an upper limit on the optical eclipse depth of $\lesssim$300 ppm which is not sufficient to distinguish between the scenarios presented here.

The bottom panel of Figure \ref{fig:emission} compares the vertical mixing ratio profiles (equilibrium chemistry) under the different scenarios as well as with the free retrieval results in Section \ref{aura} (horizontal colored lines).  In general, self-consistent assumptions and equilibrium chemistry struggle to produce the abundance patterns derived from the free retrieval.

\section{Discussion}
\label{discussion}

\subsection{KELT-11b's Unusual Water Feature}
\label{weirdness}

As discussed above, the \HST~transmission spectrum of KELT-11b appears to have revealed a water feature with an unusual shape compared to a ``normal'' water absorption signal. We refer to this as an unusual shape because of how the transmission spectrum at the red edge ($>$1.5 $\mu$m) is relatively flat and does not slope downward like typical water absorption features observed in other exoplanets do \cite[where a downward slope would indicate a lack of additional absorption at the red edge, e.g.,][]{iyer2016,fu17}. 

With a median transit depth uncertainty of just 16 ppm, the KELT-11b transmission spectrum is one of the most precise  spectra ever measured with \HST/WFC3 \citep{knutson14, guo20, Line2016}, so we carefully considered the possibility that the shape of the transmission spectrum could be biased by instrument systematic noise. To address this concern, we reduced the data with two independent pipelines and fit the light curves with three different systematics models (Section \ref{hst_spec_fits}). The measured transit depths were typically consistent to better than $1\sigma$ for all methods (see Figure\,\ref{fig:hst_model_comparison}). Despite the consistency between different light curve fits, this measurement does push the limit of precision ever achieved with WFC3, and repeat measurements would help confirm the shape of the transmission spectrum.

We also considered several astrophysical effects that could have impacted the shape of the \HST/WFC3 transmission spectrum. First, the only star we see in the \HST/WFC3~direct image is KELT-11. \citet{pepper2017} found no evidence for background/nearby companion stars at separations of 1.5–-4 arcsec that could have contaminated our transmission spectrum. $Gaia$ DR2 identified no stars within 20 arcsec of KELT-11 and no stars brighter than $Gaia$ $G$ mag $\sim$ 19 within 45 arcsec \citep{gaia2018}.

Second, the optical ground-based light curves we collected in April 2018 were used to investigate any stellar activity that may have been occurring around the time of the \HST~transit or \Spitzer~eclipse observations of KELT-11b. While of much lower precision than the \HST~and \Spitzer~data, the ground-based data (Figures \ref{fig:all_ground} and \ref{fig:each_ground}) show KELT-11's flux is stable overall. There are no obvious signs of spots or flares that could have adversely impacted the \HST~transmission spectrum or \Spitzer~eclipse measurement. \citet{pepper2017} estimate a rotational period for the star of 52 days, supporting our evidence that the stellar flux was stable over the $\sim$7 days between the \HST~transit and \Spitzer~eclipse observations collected in 2018. In addition, the $\sim$27 day light curve from \TESS~collected in February-March 2019 does not appear to display coherent, periodic variability. We conclude that the \HST~transit data (and the \Spitzer~eclipse data) were likely not collected in a time of anomalous stellar activity that could have in turn impacted the atmospheric signals we find here.

Third, KELT-11 is a slow rotator [$v\sin(i) = 2$ km/s; \citealt{pepper2017}]. This means KELT-11b should not be impacted by the phenomenon known as ``gravity-darkened seasons,'' which arise when a rapidly-rotating star (i.e. an oblate spheroid) has a varying temperature across its surface and as a consequence a planet receives a varying amount of flux throughout its orbit \citep{ahlers2016}. However, KELT-11 is part of the ``Retired A-star'' class meaning it is not implausible that the system is misaligned, since gas giants around high-mass stars are frequently spin-orbit misaligned \citep{winn2010ApJ...718L.145W}. In this scenario, KELT-11 could be a rapid rotator that we are seeing pole-on [which would result in a small $v\sin(i)$]. To test this, we performed a separate fit to the optical \TESS~transit light curve following the methods of \citet{ahlers2020} that invoke a gravity-darkened light curve model. We only fit the optical data here, because the effects of gravity darkening are chromatic with effects typically increasing toward the ultraviolet. In our fit, we find that the gravity-darkening exponent is consistent with zero (0.10$\pm$0.1), indicating that gravity-darkened seasons is not a likely scenario. We also measure a transit depth that is consistent with our other analyses of the \TESS~data.

Barring unknown instrumental and astrophysical effects at the time of the \HST~transit observations, we conclude that the \HST/WFC3 transmission spectrum of KELT-11b we present here is truly unusual in shape. Interestingly, the shape of KELT-11b's transmission spectrum is reminiscent of the transmission spectrum measured for WASP-63b, which displayed an apparent ``bump'' around 1.55$\mu$m \citep{kilpatrick2018}. In that case, the presence of HCN at super-solar abundances was considered as the potential cause of the shape of the spectrum of WASP-63b. As discussed by \citet{kilpatrick2018}, such a high abundance of HCN requires disequilibrium chemistry processes, and further observations are needed to either confirm or exclude the presence of HCN in the planet atmosphere. We come to a similar conclusion here regarding the apparent HCN feature in KELT-11b's spectrum, which we discuss further in the following sections.

\subsection{Interpreting the Composition of KELT-11b's Atmosphere}
\label{composition}

Perhaps unsurprisingly, with the unusual water absorption feature seen in KELT-11b's atmosphere, we have obtained some unusual results when performing standard analysis and retrievals on the transmission spectrum. Even though the self-consistent models using the CHIMERA tool are technically rejected by the data, we find they consistently produce an extremely low metallicity of $[M/H] \lesssim -2$ and an upper limit on the C/O ratio. In the free retrievals with the AURA tool, which we find provide more adequate fits to the data overall (owing to the larger number of free parameters), we clearly detect H$_2$O at high significance in all cases ($>$3.1$\sigma$) with an abundance that corresponds to $\lesssim$0.1$\times$ solar. We also conservatively detect HCN, which helps explain the shape of the red end of the \HST~transmission spectrum. In the fiducial model of the combined \TESS+\HST+\Spitzer~transmission spectrum, we find that AlO is preferred at 2$\sigma$; however, this is the only case where AlO was found at any significance. There is also a hint of TiO, particularly when modeling the \HST~transmission spectrum alone. A common conclusion from the AURA and CHIMERA retrievals is that KELT-11b appears to have a (very) sub-solar atmospheric metallicity, with a low-abundance water feature sitting on top of the H$_2$-H$_2$ and H$_2$-He collision induced absorption.

While HCN, AlO, and TiO are typically considered weakly detected ($<$3$\sigma$), when included, their abundances (and precisions) (Table \ref{tab:retrievals}) are at odds with predictions from equilibrium chemistry. Figure \ref{fig:emission} (bottom panel) compares representative solar (solid) and sub-solar (dashed) self-consistent equilibrium chemistry mixing ratio profiles to the Free 1 (thick horizontal lines) and Free 5 (thin horizontal lines) constraints.   From this, it is apparent that the free retrieval abundance patterns are hard pressed to match equilibrium expectations.  

Disequilibrium chemistry (e.g., vertical mixing and/or photochemistry) would also struggle to explain such patterns.  For instance, the Free 1 HCN constraint is about 1-2 orders of magnitude larger than even the deepest value (in an optimistic scenario of strong vertical mixing) of the predicted solar HCN abundance. Even in the presence of vertical mixing, a flavor of photo/ion-chemical enhancement would be required; however, typical photochemical models of planets more favorable for HCN formation [e.g., the cooler HD 209458b; \cite{moses2011}], struggle to produce detectable enhancements ($<10^{-8}$).  The presence of TiO and AlO are also unexpected given the equilibrium TP-profile.  At solar, Al has rained out into Al$_2$O$_3$ and Ti into CaTiO$_3$.  Though kinetic inhibition could perhaps play a role in halting this condensate sequence, this would be at odds with the spectral changes observed in the brown dwarf sequence \citep[e.g.,][]{Kirkpatrick2005} and kinetic cloud models \citep{helling2008A&A...485..547H,helling2019A&A...626A.133H}.  

The Free 5 (and 6) scenarios that implement an isothermal/simplified model, however, are more consistent with the low metallicity self-consistent solution, with the 1$\sigma$ lower HCN bound overlapping with HCN abundances plausibly enhanced through vertical mixing (though the precise constraint on TiO is still $\sim$2 orders of magnitude larger than predicted).\footnote{It is interesting to note that TiO persists in the low metallicity scenario as there is not enough combined Ca and Ti to form the condensates that deplete gas phase TiO.  } However, we emphasize that there are challenges in deriving abundances using simplified models. By assuming isotherms and clear atmospheres in our simplified models and not considering the impact of (inhomogeneous) clouds, we may be biasing our results and hence obtaining inaccurate and potentially low abundances \citep[e.g.,][]{Line2016,MacDonald2017,Welbanks2019b}.

\subsection{Is a Sub-Solar Metallicity Physically Plausible?}
One of the most intriguing results from the atmospheric modeling is the inference of a sub-solar water abundance for KELT-11b. In the case of the chemical equilibrium models, the atmosphere is inferred to be strongly sub-solar in composition ([Fe/H] $=\lesssim -2$). Even for the more flexible AURA retrieval, which allows non-equilibrium abundances, the inferred water abundance is still lower than the abundance expected for solar composition (0.001--0.7$\times$ solar over a range of model assumptions).

A sub-solar water abundance in a sub-Saturn is surprising from a planet formation standpoint. Formation models predict an atmospheric metal enrichment for sub-Saturns in the range of $10-100\times$ solar composition, regardless of whether the planets form interior or exterior to the water ice line \citep{fortney13,mordasini16}. In addition, interior structure models based on the observed masses and radii of gas giant exoplanets also suggest a moderate metal enrichment of $\sim10\times$ solar for planets in the sub-Saturn mass range \citep{thorngren16}. The retrieval results for KELT-11b suggest a sub-solar water abundance ($0.01 - 0.1\times$ solar), which is several orders of magnitude lower than expected. 

A potential explanation for such a low metallicity could be the formation of the planet far out in the disk beyond the CO snow line where the gas is depleted of oxygen-rich volatiles \citep{Oberg2011} and migrating inward by disk-free mechanisms, as has been proposed for some hot Jupiters \citep{Madhusudhan2014b}. Similarly, the possibility of the volatiles locked up in the core as the planet forms via pebble accretion \citep{Madhusudhan2017} or the enhancement of other volatiles in the gas relative to oxygen through pebble drift \citep{Oberg2016,Booth2017} may also contribute to the observed abundances. Testing these different scenarios require precise abundance measurements for other species, such as CO, which would be possible with future observations with the James Webb Space Telescope (\JWST). 

As an additional note, it may be that atmospheric metallicity is not representative of the bulk of the H/He envelope. A recent study of Jupiter with $Juno$ and $Galileo$ mission data invokes an inward-decreasing heavy element enrichment \citep{debras2019ApJ...872..100D}. KELT-11b could similarly have significant composition gradients in the interior, which would further complicate the interpretation of the atmospheric metallicity. 

Alternatively, it may be possible that the low water abundance measurement is spurious, either the result of unknown systematics in the data or incomplete modeling of the atmospheric chemistry. The data are pushing the limit of measurement precision for WFC3, with typical precision on the transit depths of $<20$ ppm. As discussed above, there could be an unknown, wavelength-dependent systematic in the data introducing an unphysical shape in the transmission spectrum. Another possibility is that some of the simplifying assumptions in the atmospheric retrieval are invalid at this level of precision. For example,  one simplifying assumption is that the planet's atmosphere is 1D. In reality, there may be temperature-related differences between the morning and evening terminator here \citep[e.g.,][]{kempton2017}. Retrieved atmospheric properties can be systematically biased by the day-night temperature and/or composition gradient in a planet's atmosphere \citep{caldas2019A&A...623A.161C,pluriel2020A&A...636A..66P} or to different morning-evening terminator compositions \citep{macdonald2020arXiv200311548M}. Measurements of anomalous temperatures and unusual species in exoplanet atmospheres may therefore be a result of 1D assumptions \citep[e.g.,][]{macdonald2020arXiv200311548M}. From the \Spitzer~eclipse at 4.5 $\mu$m (Figure \ref{fig:emission}), we find that KELT-11b has a dayside flux consistent with full atmospheric heat redistribution and thus may not have a significant day-to-night temperature gradient. Since this is based on a single data point, additional data are required to strengthen our understanding of the day-night circulation on KELT-11b and determine if non-1D effects have influenced the retrieved temperature and composition from KELT-11b's transmission spectrum.

\subsection{Comparison to Other Inflated Sub-Saturns}
\label{sub_saturns}

Other inflated sub-Saturns that have had their atmospheres characterized in detail to date include WASP-39b \citep{faedi2011} and WASP-107b \citep{anderson2017}. These planets have masses between 0.1 and 0.3 $M_J$ and are comparably inflated to KELT-11b, but they probe a different temperature and metallicity space than KELT-11b. Compared to KELT-11b's high equilibrium temperature of $\sim$1712~K \citep{pepper2017}, WASP-39b has an equilibrium temperature of $\sim$1030~K, and WASP-107b is significantly cooler with an equilibrium temperature of $\sim$770~K. KELT-11b orbits a slightly-evolved sub-giant star with a metal-rich host ([Fe/H] = 0.17), while WASP-39b orbits a main-sequence late G-type star with a poor metallicity of [Fe/H] = --0.12. WASP-107b orbits a K star with a solar-like metallicity of [Fe/H] = 0.02.

\begin{figure}[h!]
\begin{center}
\includegraphics[scale=0.56,angle=0]{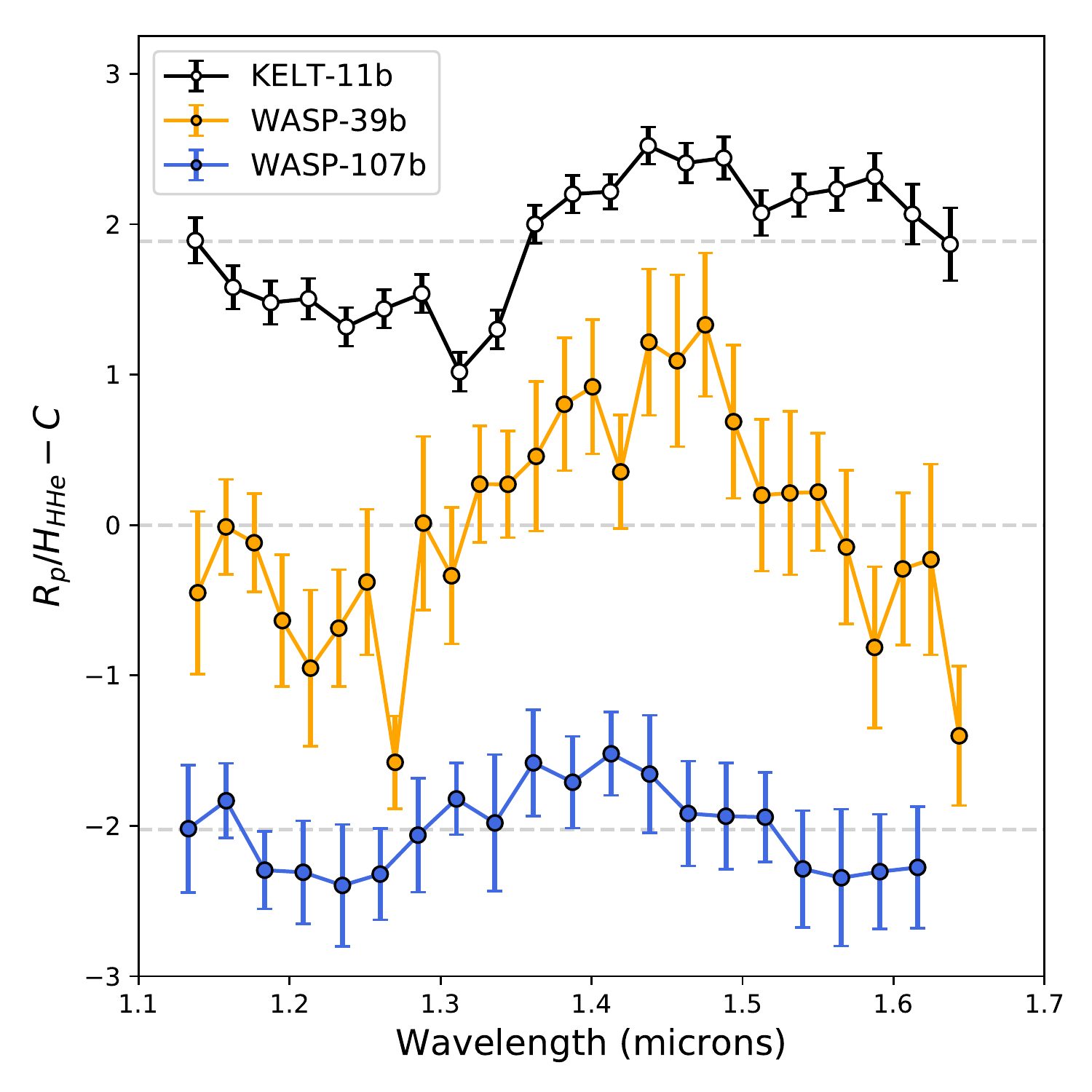}
\caption{\HST/WFC3 transmission spectra normalized to the atmospheric scale height ($H_{HHe}$) for three inflated sub-Saturns. As in \citet{crossfield2017AJ....154..261C}, we assume a H-dominated atmosphere with a mean molecular weight $\mu$ = 2.3 g mol$^{-1}$. Data for KELT-11b comes from this work, WASP-39b comes from \citet{wakeford2018}, and WASP-107b comes from \citet{kreidberg2018}. Each transmission spectrum has been offset arbitrarily by a constant $C$ for display purposes. The mean of each (offset) transmission spectrum is shown in the light gray dashed lined to guide the eye. The deviation of the KELT-11b transmission spectrum at the red edge ($>$1.5 $\mu$m) from a ``normal'' water absorption feature is clear compared to the shape of the WASP-39b and WASP-107b transmission spectra. 
\label{fig:compare_spectra}}
\end{center}
\end{figure}

\HST/WFC3 observations have revealed significant H$_2$O absorption features in the atmosphere of all these inflated sub-Saturns. Figure \ref{fig:compare_spectra} compares the KELT-11b transmission spectrum with the published WASP-39b and WASP-107b transmission spectra from \citet{wakeford2018} and \citet{kreidberg2018}, respectively. As discussed above, the shape of the KELT-11b transmission spectrum clearly deviates from the shape of a ``normal'' H$_2$O absorption feature at the red edge ($>$1.5 $\mu$m). In comparison, the spectra for WASP-39b and WASP-107b are relatively similar in shape and display the typical/expected downward trend of a H$_2$O feature at the red edge. The atmospheric metallicity of WASP-39b is estimated to be super-solar \citep{wakeford2018,kirk2019AJ....158..144K,welbanks2019ApJ...887L..20W} and for WASP-107b it is estimated to be $<$30$\times$ solar \citep{kreidberg2018}, although we note that the full range of metallicity estimates dip into sub-solar abundances \citep{welbanks2019ApJ...887L..20W}. Interior structure models for WASP-39b indicate a maximum atmospheric metallicity of $50\times$ solar, further complicating the picture \citep{thorngren19}. In this work, we performed multiple retrievals using two different tools and find that KELT-11b likely has a (very) sub-solar atmospheric metallicity (nominally 0.01--0.1$\times$ solar but 0.001--0.7$\times$ solar over a range of model assumptions). This is surprising in its own right but also given that other studies have revealed inflated sub-Saturns seem to lean towards having metal-rich atmospheres (or having only slightly sub-solar atmospheres at a minimum). This comparative activity serves to emphasize the importance of performing multiple, complementary analyses of exoplanet atmospheric data.

With a small sample of well-studied inflated sub-Saturns, it is not yet possible to look for significant trends in atmospheric characteristics with different planet or stellar properties. Theoretical predictions have been made for increasing metallicity for decreasing planet mass \citep{fortney13,mordasini16}. There is also some observational evidence that exoplanet atmospheric metallicity (based on H$_2$O) and planet mass may be correlated, with H$_2$O abundances increasing with decreasing exoplanet mass \citep{kreidberg14b,pinhas2019,welbanks2019ApJ...887L..20W}. However, precise chemical abundance determinations require a wide spectral coverage, including both optical and infrared spectra, and robust modeling and retrieval approaches.  For example, considering semi-analytic models and \HST/WFC3 data alone leads to a large scatter in the retrieved abundances with no noticeable trend (e.g., \citealp{fisher2018MNRAS.481.4698F} but c.f. \citealp{Welbanks2019a}). Therefore, additional observational data are needed to fully explore the mass-atmospheric metallicity relationship for exoplanets. 

If planet mass does not explain the differences in the atmospheric metallicity for this population of planets, other lever arms need to be explored. Host star metallicity is an intriguing option to search for correlations with atmospheric metallicity; however, we note that \citet{teske2019AJ....158..239T} found no clear correlation between stellar and planet (residual) metallicity when looking at a sample of 24 planets. This is consistent with what we see here: WASP-39b has either a metal-rich atmosphere or a slightly metal-poor atmosphere yet has a metal-poor host, and WASP-107b has a slightly metal-rich atmosphere and a solar-metallicity host. KELT-11b has a metal-rich host and might have a (significantly) metal-poor atmosphere.

Despite potential differences in the atmospheric metallicity, one common feature is that all three of the inflated sub-Saturns discussed here have significant water absorption features that are not completely hidden by clouds or hazes. Additional observations of these systems can therefore provide further insight into which planet or stellar properties play a key role in defining the constituents of the atmospheres of these inflated sub-Saturns. For instance, \citet{zak2019AJ....158..120Z} found no sign of Na absorption in KELT-11b's hot atmosphere but Na and K have been identified in the cooler atmosphere of WASP-39b \citep{wakeford2018,kirk2019AJ....158..144K}. He has been found to be escaping from the much cooler WASP-107b atmosphere around 1.1 $\mu$m \citep{spake2018,allart2019,kirk2020AJ....159..115K}, and \citet{kreidberg2018} also found that WASP-107b may be depleted in CH$_4$ relative to solar abundances. With their feature-rich atmospheres, inflated sub-Saturns are excellent laboratories for continued atmospheric characterization efforts.     

\section{Summary}
\label{summary}

Using a suite of precise \TESS~optical, \HST~near-infrared, and \Spitzer~infrared data, we have provided a detailed glimpse into the atmospheric properties of the inflated sub-Saturn KELT-11b. Our key findings are summarized below.
\begin{itemize}
    \item The \HST/WFC3 transmission spectrum of KELT-11b displays a low-amplitude absorption feature ($\sim$1 atmospheric scale height) with a shape that is unusual compared to the H$_2$O absorption features typically seen for other giant exoplanets (Figure \ref{fig:compare_spectra}).
    \item Self-consistent models of the \TESS+\HST+\Spitzer \\ combined transmission spectrum as well as the \HST~transmission spectrum alone using the CHIMERA retrieval tool produce an extremely low metallicity ($[M/H] \lesssim -2$), but all of the best fits are considered strongly rejected by the data and should therefore be interpreted with caution.
    \item Free retrieval models using the AURA retrieval tool generally provide better fits to both the combined transmission spectrum and the \HST~transmission spectrum alone. The inferred metallicity is not as extreme  ($0.001 - 0.7\times$ solar for a range of model assumptions), but additional absorbers that are far out of chemical equilibrium (e.g., HCN) are needed to explain the shape of the \HST~transmission spectrum. There is tentative evidence for other absorbing species (TiO and AlO), but the significance of these detections is sensitive to model assumptions.
    \item The dayside flux measured from the \Spitzer~eclipse is suggestive of full heat redistribution from KELT-11b's dayside to nightside, although clouds on the dayside of the planet could affect our interpretation. 
\end{itemize}
Additional observations are clearly needed to disentangle and validate these findings. \HST~observations at shorter wavelengths $<$1.1 $\mu$m would enable further analysis of the atmosphere of this intriguing exoplanet and constrain key cloud/haze parameters and the presence of alkali metals in particular. A transit of KELT-11b with the \HST/WFC3 G102 filter (0.9--1.1 $\mu$m) is currently scheduled for early 2021 (\HST~Program GO 15926: PI: K. Col\'on). In addition, the CHaracterising ExOPlanets Satellite (\CHEOPS) mission recently observed a transit of KELT-11b as part of its commissioning activities, producing a precise optical light curve over the \CHEOPS~bandpass of 0.4--1.1 $\mu$m \citep{benz2020}. A planet-star radius ratio of 0.0463$\pm$0.0003 was measured by \CHEOPS~\citep{benz2020}, which is consistent with our measured radius ratio of 0.04644$\pm$0.00065 from \TESS~(Table \ref{tab:tess_params}). With lower scatter compared to \TESS~observations ($\sim$200ppm compared to $\sim$500ppm), the \CHEOPS~observations will be useful for future analyses of the KELT-11 system. In particular, \citet{benz2020} report seeing stellar variability in the $\sim$14-hour-long \CHEOPS~light curve of KELT-11 with an amplitude of approximately 200ppm that is correlated over 30-minute to 4-hour timescales. \citet{benz2020} attribute this variability to effects of stellar granulation. Future analyses of this system should explore the impact of this potential stellar variability on transmission spectroscopy measurements. Furthermore, the upcoming infrared \JWST~will provide us with exquisitely precise data that are expected to have a typical precision on the order of $<$20 ppm for a target like KELT-11b. With \JWST, we will have the opportunity to probe the C/O ratio in detail for this inflated planet that will build on the current work and ultimately help us better understand the potential formation pathways for inflated sub-Saturns. 
As a final note, we caution that high-precision spectra of transiting exoplanet atmospheres -- like the \HST/WFC3 transmission spectrum of KELT-11b presented here that has a median uncertainty of 16 ppm -- may reveal significant atmospheric features, but they may also present new challenges for atmospheric retrieval models. We also may be venturing into a new realm of instrumental systematics that need to be considered in the reduction and analysis of atmospheric data at these levels of precision. We encourage the continued use of multiple complementary analyses of high-precision exoplanet atmosphere spectra to support any findings, especially in the coming era of \JWST~and as the exoplanet community pushes towards the atmospheric characterization of high-profile, potentially rocky planets in the habitable zones of their stars.

\acknowledgments 
We thank the referee for their thorough report, which helped us to improve this paper. We also thank Hannah Wakeford, Brian Kilpatrick, and Romain Allart for their helpful feedback. This research is based in part on observations made with the NASA/ESA Hubble Space Telescope, obtained from the Data Archive at the Space Telescope Science Institute, which is operated by the Association of Universities for Research in Astronomy, Inc., under NASA contract NAS5-26555. These observations are associated with program \#15255. This work is based in part on observations made with the Spitzer Space Telescope, which was operated by the Jet Propulsion Laboratory, California Institute of Technology under a contract with NASA. This paper includes data collected by the TESS mission, which are publicly available from the Mikulski Archive for Space Telescopes (MAST) and produced by the Science Processing Operations Center (SPOC) at NASA Ames Research Center \citep{jenkins2016,2017ksci.rept....9J}. This research effort made use of systematic error-corrected (PDC-SAP) photometry \citep{smith2012kepler,stumpe2012kepler,stumpe2014multiscale}. Funding for the TESS mission is provided by NASA's Science Mission directorate. Resources supporting this work were provided by the NASA High-End Computing (HEC) Program through the NASA Advanced Supercomputing (NAS) Division at Ames Research Center for the production of the SPOC data products. This work also used data products from the European Space Agency (ESA) mission $Gaia$ (http://www.cosmos.esa.int/gaia), processed by the $Gaia$ Data Processing and Analysis Consortium (DPAC, http://www.cosmos.esa.int/web/gaia/dpac/consortium). Funding for the DPAC has been provided by national institutions, in particular the institutions participating in the $Gaia$ Multilateral Agreement. This work has made use of NASA’s Astrophysics Data System and the SIMBAD database operated at CDS, Strasbourg, France. This research has made use of the NASA Exoplanet Archive, which is operated by the California Institute of Technology, under contract with the National Aeronautics and Space Administration under the Exoplanet Exploration Program. 

Support for program \#15255 was provided by NASA through a grant from the Space Telescope Science Institute, which is operated by the Association of Universities for Research in Astronomy, Inc., under NASA contract NAS5-26555. K.D.C., A.M., T.B., E.D.L., J.P.A., and E.A.G. acknowledge support from the GSFC Sellers Exoplanet Environments Collaboration (SEEC), which is funded in part by the NASA Planetary Science Division’s Internal Scientist Funding Model. M.R.L. acknowledges Research Computing at Arizona State University for providing HPC resources that have contributed to the research results reported within this paper. K.G.S. acknowledges support from NASA grant 17-XRP17 2-0024. J.P.A.’s research was supported by an appointment to the NASA Postdoctoral Program at the NASA Goddard Space Flight center, administered by Universities Space Research Association under contract with NASA. P.P. acknowledges support from NASA (award 16-APROBES16-0020), the National Science Foundation (Astronomy and Astrophysics grant 1716202), the Mt Cuba Astronomical Foundation, and George Mason University instructional equipment funds. K.K.M. acknowledges the support of the Theodore Dunham, Jr. Fund for Astronomical Research and the NASA Massachusetts Space Grant consortium. E.A.G. thanks the LSSTC Data Science Fellowship Program, which is funded by LSSTC, NSF Cybertraining Grant \#1829740, the Brinson Foundation, and the Moore Foundation; her participation in the program has benefited this work.

\facilities{Exoplanet Archive, HST, Spitzer, TESS}
\software{AstroImageJ \citep{collins2017}, Astropy \citep{exoplanet:astropy13}, celerite \citep{exoplanet:foremanmackey17,exoplanet:foremanmackey18,exoplanet:astropy18}, batman \citep{kreidberg15}, emcee \citep{foremanmackey13}, exoplanet \citep{exoplanet:exoplanet}, Lightkurve \citep{Lightkurve}, matplotlib \citep{matplotlib}, PyMC3 \citep{exoplanet:pymc3}, Theano \citep{exoplanet:theano}, starry \citep{exoplanet:luger18}, TAPIR \citep{jensen2013ascl.soft06007J}}

\clearpage
\bibliography{bibliography}

\listofchanges

\end{document}